\documentclass[showpacs,showkeys,preprint,amsmath,amssymb,nofootinbib, prd, aps, 11pt]{revtex4}
\usepackage[T1]{fontenc}
\usepackage[latin1]{inputenc}
\usepackage{slashed}
\usepackage{mathtools}

\usepackage{mathrsfs}
\usepackage{color}
\usepackage{framed}
\usepackage{graphicx}

\definecolor{shadecolor}{rgb}{1,0.94,0.72}

\usepackage[dvipsnames]{xcolor}
\usepackage{changes}
\setremarkmarkup{[\textsc{#2}]}

\newcommand{\Vek}[1]{\mbox{\boldmath$#1$\unboldmath}}

\newcommand{\eq}[1]{eq.~(\ref{#1})}

\newcommand{\zr}[1]{\mbox{\hspace*{#1em}}}
\newcommand{\ID}{\mbox{{\sf 1}\zr{-0.16}\rule{0.04em}{1.55ex}\zr{0.1}}}

\begin{document}


\title{Quantum stabilization of a hedgehog type of cosmic string}
\author{M.~Quandt}
\email{markus.quandt@uni-tuebingen.de}
\affiliation{Institute for Theoretical Physics, University of T\"ubingen, D-72076 T\"ubingen, Germany}
\author{N.~Graham}
\email{ngraham@middlebury.edu}
\affiliation{Department of Physics, Middlebury College, Middlebury, Vermont 05753, USA}
\author{H.~Weigel}
\email{weigel@sun.ac.za}
\affiliation{Institute for Theoretical Physics, Stellenbosch University, Matieland 7602, South Africa}


\begin{abstract}
Within a slightly simplified version of the electroweak standard model we investigate 
the stabilization of cosmic strings by fermion quantum fluctuations. Previous studies 
of quantum energies considered variants of the Nielsen-Olesen profile embedded in the 
electroweak gauge group and showed that configurations are favored for which the Higgs 
vacuum expectation value drops near the string core and the gauge field is suppressed.  
This work found that the strongest binding was obtained from strings that differ 
significantly from Nielsen-Olesen configurations, deforming essentially only 
the Higgs field in order to generate a strong attraction without inducing 
large gradients.  Extending this analysis, we consider the leading quantum correction 
to the energy per unit length of a hedgehog type string, which, in contrast to the 
Nielsen-Olesen configuration, contains a pseudoscalar field. To employ the spectral 
method we develop the scattering and bound state problems for fermions in 
the background of a hedgehog string. Explicit occupation of bound state levels leads 
to strings that carry the quantum numbers of the bound fermions. We discuss the 
parameter space for which stable, hedgehog type cosmic strings emerge and reflect 
on phenomenological consequences of these findings.
\end{abstract}


\pacs{03.50.De, 03.65.Nk, 11.15.Kc, 11.80.Gw}
\keywords{} 
\maketitle


\section{Introduction and motivation}
The electroweak standard model and many of its extensions have the potential to 
support string--like configurations. These field configurations are the particle 
physics analogs of vortices or magnetic flux tubes in condensed matter physics. 
They are usually called \emph{cosmic strings} to distinguish them from the 
fundamental variables in string theory, and also to indicate that they typically 
stretch over cosmic length scales. In the context of the standard
model they are also called $Z$ (or $W$)
strings~\cite{Vachaspati:1992fi,Achucarro:1999it,Nambu:1977ag}
to illustrate that are composed of massive gauge fields.

The topology of string--like configurations is described by
the first homotopy group $\Pi_1(\mathscr{M})$, where $\mathscr{M}$ is the
manifold of vacuum field configurations far away from the string.
In typical electroweak-type models, a Higgs condensate breaks an
initial gauge group $G$ down to some subgroup $H$, so that
$\mathscr{M} \simeq G/H$. Topologically stable strings are therefore ruled out
in the electroweak standard model $SU(2) \times U(1) \to U(1)$ because
$G/H$ is simply connected. Nevertheless, one could envision a GUT  
and/or supersymmetric extension in which a simply connected group $G$ 
breaks down to the electroweak $SU(2) \times U(1)$ at a much higher scale, so that
$\Pi_1(G / (SU(2)\times U(1)))$ is nontrivial and strings would be 
topologically stable in such GUTs. These strings would have enormous 
energy densities, so that they could be seen by direct observation 
using gravitational lensing \cite{Kibble:1976sj,Hindmarsh:1994re}
or by signatures in the cosmic microwave background \cite{Shellard}.
Moreover, a network of  such strings is a candidate for the
dark energy required to explain the recently observed cosmic
acceleration \cite{Perlmutter:1998np,Riess:1998cb}.

The absence of topological stability does not imply that the $Z$ strings at 
the electroweak scale are unstable or irrelevant for particle physics. 
While their direct gravitational effects are small, $Z$--strings can 
still be relevant for cosmology 
at a sub--dominant level \cite{Copeland:2009ga,Achucarro:2008fn}.
Their most interesting consequences originate, however, from their
coupling to the standard model fields. $Z$--strings provide a source
for primordial magnetic fields \cite{Nambu:1977ag}
and they also offer a scenario for baryogenesis with a second order
phase transition~\cite{Brandenberger:1992ys}.
In contrast, a strong first order transition as required by the usual bubble
nucleation scenario is unlikely in the electroweak standard model
\cite{EWPhase} without non-standard additions such as supersymmetry or 
higher--dimensional operators \cite{Grojean:2004xa}. When a string
changes its shape baryon number violation may occur, but for baryogenesis 
to prevail after the string has 
disappeared an additional process, {\it e.g} via a sphaleron transition, is 
required \cite{Sato:1995ea}. Also de-linking closed $Z$--strings change their 
helicity (Chern-Simions number)  which in turn induces baryon number 
violation \cite{Dziarmaga:1994sg}. Yet, the baryon number generation from 
$Z$ strings is not sufficient to explain the observed abundance \cite{Cline:1998rc}.

However, such effects are only viable if the cosmic strings are at least meta-stable, 
such that they live long enough to have a cosmological impact. Classically, the 
energy required to wind up an electroweak string of astrophysical length scales 
is huge, but it may eventually be overcome by quantum effects induced by the 
coupling to the remaining fields. In this respect, the most important contributions 
are expected to come from heavy fermions, since their quantum energy dominates 
in the limit $N_C \to \infty$, where $N_C$ is the number of QCD colors or other 
internal degrees of freedom.  Heavier fermions are expected to provide more binding 
since the energy gain per fermion charge is higher and their Yukawa coupling to 
the string is larger; a similar conclusion can also be drawn from decoupling 
arguments \cite{D'Hoker:1984ph}. Generally, the string background 
deforms the Dirac spectrum and typically leads to the formation of either an 
exact or near zero mode \cite{Naculich:1995cb}, so that fermions can substantially 
lower their energy by binding to the string, which may eventually overcome the 
classical energy cost of building the string. For consistency,
however, one must include \emph{all} contributions which have the same formal loop 
order; in particular, this means that the deformation of the continuous
part of the spectrum (the vacuum polarization energy) must be taken
into account as well. 

A number of previous studies have investigated quantum properties of string
configurations. Naculich~\cite{Naculich:1995cb} has shown that in the limit of 
weak coupling, fermion fluctuations destabilize the string. The quantum properties 
of $Z$--strings have also been connected to non--perturbative 
anomalies~\cite{Klinkhamer:2003hz}. The emergence or absence of exact neutrino 
zero modes in a $Z$--string background and the possible consequences for the 
string topology were investigated in Ref.~\cite{Stojkovic}. A first attempt at 
a full calculation of the fermionic quantum corrections to the $Z$--string 
energy was carried out in ref.~\cite{Groves:1999ks}. In that work, the authors 
could not compare the cosmic string to the perturbative vacuum because of the 
non-trivial winding of the string background at spatial infinity. Methods to 
overcome that technical problem were developed a decade 
later \cite{Weigel:2010pf,Weigel:2016ncb}. The first comprehensive calculation of 
the fermionic vacuum polarization energy of the Abelian Nielsen--Olesen 
vortex~\cite{Nielsen:1973cs} has been estimated in ref.~\cite{Bordag:2003at}, 
where subtractions were carried out in the heat--kernel expansion, which is 
not easily connected with the standard perturbation counterterms. Quantum energies of 
bosonic fluctuations in string backgrounds were calculated in ref.~\cite{Baacke:2008sq}. 
Finally, the dynamical fields coupled to the string can also result in (Abelian or 
non--Abelian) currents running along the core of the string. The time evolution 
of such structured strings was studied in ref.~\cite{Lilley:2010av}, where the 
current was induced by the coupling to an extra scalar field.

Mathematically, the problem of computing the leading quantum energy of a string background 
amounts to the computation of the determinant for the Dirac operator within this background. 
Previously, we have employed the \emph{spectral method} to study the quantum energy of a 
special type of cosmic string in a reduced version of the standard model 
\cite{Weigel:2010zk,Graham:2011fw}.  
Even though we allowed for a non-trivial gauge-field structure in the cosmic string 
background, the findings from Ref.~\cite{Graham:2011fw} indicate that the preferred 
string configuration has very little gauge field admixture. Instead, it reduces to a 
narrow ditch carved in the Higgs condensate. In the present study, we will follow 
up on the observation that the Higgs field is the dominating factor but
consider a different mechanism, inspired by topological solitons, in order to
produce attraction in the scalar potential for the fermions and thus 
generate binding for the fermions. In many non-linear bosonic models such as the 
Skyrme model \cite{Skyrme:1961vq}, the classical solutions of the field equations 
(i.e.~the static configurations with minimal energy) that support an extended region of 
suppressed condensate have a characteristic \emph{hedgehog} structure. When coupled
to fermions, as {\it e.g.} in the Nambu-Jona-Lasinio soliton model \cite{Alkofer:1994ph},
the hedgehog configuration produces strong binding even when the magnitude of the scalar 
component of the Higgs field is homogeneous. Hence this configuration may contribute a 
significantly lower classical energy for the same gain from the fermion quantum energy. We 
formulate the two dimensional analog of the hedgehog configuration in the plane 
perpendicular to the string and extend it uniformly  along the string.
We couple fermions to this configuration and compute the resulting spectrum.  After 
proper renormalization this spectrum yields the vacuum polarization energy, the numerical 
simulation of which will determine whether or not such hedgehog structures 
with shallow scalar Higgs components are energetically favored.

This paper is organized as follows: In the next section we describe our model and 
introduce the hedgehog type of string configuration. In section~\ref{sec:tools}, we 
adapt methods from Refs.~\cite{Weigel:2010zk, Graham:2011fw} to compute the fermion 
vacuum polarization energy to this hedgehog configuration. This calculation requires
finding the Jost determinant from scattering data and, via the Born series, combining 
the spectral method with explicit calculations of low-order Feynman diagrams. Then the 
quantum energy can be renormalized with conventional ($\overline{MS}$ or \emph{on-shell}) 
schemes, allowing for the  model parameters to be specified from phenomenological data.
In that section we also explain how the string is equipped with charge.

In section \ref{sec:results}, we present our results for both neutral and charged strings. 
We also relax our string background profile to allow for a more shallow suppression of 
the scalar component of the Higgs background, which has smaller classical costs but also 
tends to bind the majority of fermions less deeply. In our variational approach the
optimal configuration for each given charge is selected from several hundred distinct 
string profiles, and the minimal fermion mass required for a stable configuration is estimated. 
In section \ref{sec:summary}, we briefly summarize and discuss our findings and comment 
on possible consequences for cosmology or particle physics.  The technicalities of the 
scattering problem and the renormalization procedure are described in detail in appendices.

\section{Cosmic strings in a simplified electroweak model}
\label{sec:model}

A cosmic string is a line-like soliton within electroweak or grand unified type theories.
If the gauge group is simply connected ($ \pi_1(G) = \emptyset $), there is no topological 
argument in favor of (classical) stability, and the string must be stabilized dynamically,
e.g.~by reducing its energy via quantum fluctuations. In Ref.~\cite{Graham:2011fw}, we have 
studied this scenario in a slightly simplified version of the $SU(2)$ electroweak theory,
\begin{align}
\mathcal{L} = - \frac{1}{2}\,\mathrm{tr}\,\big( G^{\mu\nu}\,G_{\mu\nu}\Big) &+ \frac{1}{2}\,
\mathrm{tr}\,\big(D_\mu \Phi\big)^\dagger\,\big(D^\mu \Phi\big) - \frac{\lambda}{2}\,
\mathrm{tr}\,\big(\Phi^\dagger\Phi - v^2\big)^2 + 
\nonumber \\[2mm]
{} &+ i \,\bar{\Psi}\,\big(\mathsf{P}_L\,D\!\!\!\!/ + \mathsf{P}_R\,\partial\!\!\!/\big)\Psi 
- f\,\bar{\Psi}\,\big(\Phi \,\mathsf{P}_R + \Phi^\dagger\,\mathsf{P}_L\big)\,\Psi\,.
\label{model}
\end{align}
Here, the first three terms describe the bosonic sector made up of weak gauge bosons $W_\mu$ 
with non-Abelian field strength $G_{\mu\nu}=\partial_\mu W_\nu-\partial_\nu W_\mu 
+ ig[W_\mu, W_\nu]$, and gauge coupling $g$ as well as the Higgs doublet $\Phi$ in 
the fundamental representation of the weak isospin group $SU(2)$. The fourth and fifth terms 
denote the fermion sector with the minimal coupling of the left-handed quarks 
to the bosonic sector. Both, the Higgs and the fermion fields couple to the gauge bosons
via the covariant derivative $D_\mu=\partial_\mu-igW_\mu$. The simplifications of 
Eq.~(\ref{model}) as compared to the standard model are: \textbf{(i)} the Weinberg angle is 
set to zero and the $U(1)$ hypercharge is discarded, \textbf{(ii)} the fermion doublet is 
taken to be degenerate in mass with inter-family fermion mixing neglected, \textbf{(iii)} 
only the heaviest quark doublet is retained, since it has the strongest coupling to the Higgs 
field; see  Ref.~\cite{Graham:2011fw} for further details on the justification 
of these assumptions.

The string configuration is translationally invariant and is infinitely extended along its 
symmetry axis. We adapt an {\it ansatz} that has the typical string-like suppression of the Higgs 
condensate in the vicinity of the symmetry axis, with no gauge field decoration, {\it i.e.} 
$W_\mu=0$. This suppression of the Higgs condensate defines the string core. In contrast to the 
Nielsen-Olesen configuration, the winding of the Higgs field around the symmetry axis decays 
asymptotically for the background that we entertain here. This requires independent profile functions 
for the charged and neutral Higgs fields in the plane perpendicular to the symmetry axis, which we 
take to be the $z$-axis with polar coordinates $r$ and $\varphi$ in the \emph{xy}-plane. 
Then the two profile functions $\rho(r)$ and $\theta(r)$, respectively called chiral radius and 
chiral angle, parameterize the Higgs field in its matrix representation via
\begin{align}
\Phi = v \,\rho(r)\,
\begin{pmatrix}
\cos \theta(r) & i e^{i \varphi}\,\sin\theta(r) \\[2mm]
ie^{-i \varphi}\,\sin \theta(r) & \cos \theta(r)
\end{pmatrix} \,,
\label{config}
\end{align}
which is related to the common doublet notation by
\begin{align}
\varphi = \begin{pmatrix} \varphi_+ \\ \varphi_0 \end{pmatrix}\qquad\quad
\Longleftrightarrow\qquad\quad
\Phi = 
\begin{pmatrix} 
 \varphi_0^\ast & \varphi_+  \\[2mm] -\varphi_+^\ast & \varphi_0      
\end{pmatrix}\,.
\end{align}
The string background can then be re-written in the form
\begin{align}
\Phi = v\,\Big[s(r) + i \,(\Vek{\tau}\cdot\hat{\Vek{r}})^\ast\,p(r)\Big] 
\label{config1}
\end{align}
where $\Vek{\tau}$ are the isospin Pauli matrices. This defines the scalar and pseudo-scalar 
profile functions 
\begin{align}
s(r) = \rho(r)\,\cos\theta(r) \qquad {\rm and}\qquad
p(r) = \rho(r)\,\sin\theta(r)\,,
\label{profiles}
\end{align}
which illuminate the relation to the Skyrme model\footnote{It should be emphasized, however, that 
the Skyrme equations are merely a motivation and the configuration (\ref{config}) is \emph{not} 
necessarily a solution of the equations of motion for the model eq.~(\ref{model}), nor is this 
necessary for the following.}, justifying the identification
of our configuration as a hedgehog background. 

The vacuum expectation value (\emph{vev}) of the Higgs doublet is at the minimum of the 
potential, i.e.
\begin{align}
\langle \| \varphi \|^2 \rangle = \langle |\varphi_0|^2 + |\varphi_+|^2 \rangle 
= \langle\,\det(\Phi)\rangle = \frac{1}{2}\mathrm{tr}\langle\Phi^\dagger\Phi\rangle
= v^2 \,.
\end{align}
(Note that our convention differs slightly from the standard one, which parameterizes the 
classical minimum as $\frac{\mu^2}{2}$.) The Yukawa coupling to the quarks gives rise to 
the quark mass $m=vf=\mu f/\sqrt{2}$. Phenomenologically, the standard Higgs scale is 
$\mu=246\,\mathrm{GeV}$, so that $v=174\,\mathrm{GeV}$. For the top quark this corresponds 
to a Yukawa coupling of 
\begin{align}
f(\text{top}) = \frac{173\,\mathrm{GeV}}{174\,\mathrm{GeV}} = 0.99\,.
\label{mtop}
\end{align}
The Higgs coupling $ \lambda $ determines the ratio of the Higgs mass and \emph{vev}. 
More precisely, our convention for the potential gives $ m_H^2 = 4 \lambda v^2 $ and hence
\begin{align}
\lambda = \frac{m_H^2}{4 \,v^2} = \frac{(125\,\mathrm{GeV})^2}{4(174\,\mathrm{GeV})^2}
= 0.129\,.
\label{mHiggs}
\end{align}
It should be stressed again that the two couplings, $f$ and $\lambda$, are dimensionless and 
independent: once the Higgs \emph{vev} is fixed, the Yukawa coupling determines the fermion 
mass, and the Higgs coupling determines the Higgs mass. In particular, $\lambda$ is completely 
obtained from properties of the Higgs field alone. It is therefore convenient to leave the 
Higgs sector fixed with $\lambda=0.129$, and vary the Yukawa coupling from its top quark
value, Eq.~(\ref{mtop}) to study the effect of different quark masses.

\medskip\noindent
The classical energy per unit length of the string configuration (\ref{config1}) is 
obtained by substituting the profiles, Eq.~(\ref{config}), into the (negative) Lagrangian,
Eq.~(\ref{model}), integrating over space, and dividing by the (infinite) length, $L_z$
of the string:
\begin{align}
\frac{\mathcal{E}_{\rm cl}}{m^2} = \frac{E_{\rm cl} / L_z}{m^2}
 &= \frac{2\pi}{f^2}\,\int_0^\infty dr\,r\,\Bigg(\frac{\rho(r)^2}{r^2}\,\sin^2 \theta(r) 
 + \rho(r)^2\,\theta'(r)^2 + \rho'(r)^2 + \frac{\lambda}{f^2}\,
 \big[1 - \rho(r)^2\big]^2 \Bigg)
 \nonumber \\[2mm]
 &= \frac{2\pi}{f^2}\,\int_0^\infty dr\,r\,\Bigg(\frac{p(r)^2}{r^2} + 
 s'(r)^2 + p'(r)^2 + \frac{\lambda}{f^2}\,\big[1 - s(r)^2 - p(r)^2\big]^2\Bigg)
 \label{Eclass}
\end{align}
Here and in the following, all dimensionful quantities are measured in appropriate units of 
the quark mass $m$: for instance, the dimensionless radial distance $r$ in eq.~(\ref{Eclass}) 
is really $ \hat{r} \equiv m r $, but we omit the hat for simplicity.

We require that the background configuration has finite classical energy (per unit length). 
At large distances from the string core, this implies that the Higgs is in its vacuum state
($\rho = 1$), and $\sin\theta = 0$ to avoid the logarithmic divergence in the first term 
under the integral in Eq.~(\ref{Eclass}). Unless $\sin\theta\to0$ as $r\to 0$, the same term 
has divergences at short distances because we want to allow $\rho(0)=\rho_0$ to take any value. 
Altogether, the requirement of finite energy enforces the following boundary conditions for 
the two profile functions in our configuration: 
\begin{align}
r \to 0 \,&:\,& 
&\rho(r) \to \rho_0\,& &\theta(r) \to \nu_0 \pi\qquad(\nu_0 \in \mathbb{Z})
\nonumber\\
& & & s(r) \to \mp \rho_0 & & p(r) \to 0
\nonumber \\[2mm]
r \to \infty \,&:\, & &\rho(r) \to 1 & & \theta(r) 
\to \nu_\infty \pi\qquad(\nu_\infty \in \mathbb{Z})
\nonumber \\
&  & & s(r) \to \pm 1 & & p(r) \to 0\,.
\label{103}
\end{align}
The integer numbers in the boundary condition for the chiral angle are
conventionally chosen as $\nu_0=-1$ and $\nu_\infty=0$, leading to the 
upper sign in the boundary values for the scalar profile $s(r)$. For most of this
study, we will assume that the Higgs condensate vanishes at the string core, 
$\rho_0 = 0$, since this leads to deeply bound fermion states located near the 
string core, which is beneficial for a possible quantum stabilization. Alternatively, 
more shallow configurations with $0 < \rho_0 < 1$ induce less binding in the 
quantum energy, but also have a smaller classical energy to overcome, so 
that an attractive net effect may emerge as motivated in the introduction. 

\section{Quantum corrections to the string energy}
\label{sec:tools}
In the limit of a large number of external quantum numbers (e.g. the number of quark colors 
$ N_c \gg 1 $), the leading quantum corrections to the classical energy of the cosmic string
originate from the fluctuations of the Dirac fermion $ \Psi $. For time-independent background 
fields, this sector is governed by the single-particle Hamiltonian
\begin{align}
H = - i \Vek{\alpha}\cdot\Vek{\nabla} \otimes \mathbf{1}_I
+ \frac{f}{2}\,\beta\,\big(\Phi + \Phi^\dagger\big)  
+ \frac{f}{2}\,\beta\gamma_5\,\big(\Phi - \Phi^\dagger\big)\,,
\label{Ham}
\end{align}
where $\boldsymbol{\alpha}$, $\beta$ and $\gamma_5$ are the usual Dirac matrices and 
$\mathbf{1}_I$ is the $(2\times 2)$ unit matrix in weak isospin space. The entire 
Hamiltonian acts on 8-component Dirac $\times$ isospin spinors.  We split the Dirac 
Hamiltonian in a free and interaction part, $ H = H_0 + H_{\rm int}$, with 
\begin{align}
H_0 &= 
\Big[ -i \Vek{\alpha}\cdot\hat{\Vek{r}}\,\partial_r - 
i \,\big[\Vek{\alpha}\cdot\hat{\Vek{\varphi}}\,
r^{-1}\,\partial_\varphi - i\,\Vek{\alpha}\cdot\hat{\Vek{z}}\,\partial_z 
 + \beta m \Big]\,\otimes \mathbf{1}_I
\label{hfree}
\\[2mm]
H_{\rm int} &= \beta\left(\frac{f}{2}\,\big[\Phi + \Phi^\dagger\big] 
- \mathbf{1}_I \right) 
+ \beta\,\gamma_5\frac{f}{2}\,\big[\Phi - \Phi^\dagger\big] 
\nonumber \\[2mm]
&= \beta \otimes \mathbf{1}_I \,\Big[s (r) - 1\Big]\,m +
i\,(\beta \gamma_5) \otimes I_\varphi\,p(r)\,m\,,
\label{hint}
\end{align}
where the Dirac and isospin matrices in the interaction are given explicitly by
\begin{align}
\beta = \begin{pmatrix} \mathbf{1} & 0 \\ 0 & -\mathbf{1} \end{pmatrix}\,,\qquad\quad
\beta\,\gamma_5 = \begin{pmatrix} 0 & \mathbf{1} \\ -\mathbf{1} & 0 \end{pmatrix}\,,\qquad\quad
I_\varphi \equiv \left(\Vek{\tau}\cdot\hat{\Vek{r}}\right)^\ast 
= \begin{pmatrix} 0 & e^{i\varphi} \\ e^{-i \varphi} & 0 \end{pmatrix}\,.
\end{align}
Form the boundary conditions, Eq.~(\ref{103}) we observe that $H_{\rm int}\to0$ as $r\to\infty$. 
This differs significantly from configurations that are variants of the Nielsen-Olesen string 
and approach a pure gauge configuration asymptotically. This difference simplifies the computation 
considerably, since no artificial gauge field is needed to map this pure gauge onto the 
trivial configuration~\cite{Weigel:2010pf}. 

We have omitted the trivial part $(-i\alpha_3 \partial_z)\otimes \mathbf{1}_I$ in $H_0$, since 
the background is translationally invariant in $z$-direction. It produces the factor 
$\sim e^{ipz}$ for the full wave functions and its contribution to the vacuum polarization energy 
is accounted for by the \emph{interface formalism} that we will introduce below.

The spectrum obtained from $H$ will always be charge conjugation invariant because 
$\left\{H,\alpha_3\right\}=0$. This invariance implies that the polarized vacuum has zero 
charge and that the biggest energy gain from a single particle level is $m$. In contrast,
the three-dimensional hedgehog does not have this symmetry, so it can carry a vacuum charge 
and can have an energy gain as big as $2m$ from a single level.

\subsection{Contributions to the quantum energy}

The energies of single particle harmonic fluctuations are altered 
by the interaction with the background. At one loop order the quantum energy 
is the renormalized sum of these energy shifts, which we compute using 
the spectral method \cite{Graham:2009zz}. In this formalism, both
isolated bound states and continuum scattering states contribute 
to the quantum or \emph{vacuum polarization} energy.
The continuum contribution can be expressed as the momentum ($k$) integral over the 
product of single particle energies $\omega=\sqrt{k^2+m^2}$ and the change in the 
density of states for that $k$, which in turn is related to the momentum
derivative of the scattering phase shift $\delta(k)$. For the string this sum 
is not sufficient because the trivial exponential factor $e^{ipz}$ changes the 
dispersion to $\omega=\sqrt{k^2+p^2+m^2}$. Potential divergences originating 
from the additional momentum integral cancel between the bound state 
and continuum contributions due to particular sum rules for scattering 
data \cite{Puff:1975zz,Graham:2001iv}. In essence the $p$ integral produces 
an additional energy factor under the $k$ integral. This is the main result 
of the interface formalism \cite{Graham:2001dy}. For calculational purposes 
the phase shift is expressed as the phase of the Jost function which
has zeros at imaginary momenta representing the bound states. Since we require
the logarithmic derivative of the Jost function, the contour integral in 
complex momentum space automatically accounts for the bound state contribution 
and the sole contribution stems from the discontinuity of the dispersion relation 
on the imaginary axis, $k=it$ for real $t\ge m$ \cite{Graham:2002xq,Graham:2009zz}. 
This produces the spectral integral 
\begin{align}
\mathcal{E}_q \sim - N_c \int_m^\infty \frac{dt}{4\pi}\,t \,u(t) = 
- N_c \int_0^\infty \frac{d\tau}{4\pi}\,\tau \,u\left(\sqrt{\tau^2 + m^2}\right)\,,
\label{evac0}
\end{align}
which is a formal result because regularization and renormalization has yet to 
be implemented. The integrand $u(t)$ has the partial wave decomposition
\begin{align}
u(t) &\equiv 2 \,u_F(t) = 2 \sum_{\ell = -1}^\infty D_\ell\, \nu_\ell(t)\,.
\label{defu}
\end{align}
In appendix \ref{app:scat} we describe in great detail the partial wave decomposition of
$\nu_\ell(t)$ and how it is obtained as the logarithm of the Jost determinant from the 
solutions to the Dirac equation. The degeneracy of the angular momentum channel 
$\ell=-1,0,1,\ldots$ is $2D_\ell=2(2-\delta_{\ell,-1})$, due to the sum over both Riemann 
sheets in the relativistic fermion dispersion relation.

As it stands, Eq.~(\ref{evac0}) is divergent and must be combined with counterterms 
to obtain a meaningful result.  First,
we note that the integral in Eq.~(\ref{evac0}) is rendered finite by subtracting 
sufficiently many leading terms of its Born series from
$\nu_\ell(t)$. As shown in appendix \ref{app:scat}, it represents a power expansion in 
the interaction, Eq.~(\ref{hint}). Once we subtract those Born terms, we need to add them 
back as expressions that are suitable for renormalization. At this point the alternative 
formulation of the vacuum polarization energy via the functional determinant
\begin{align}
\mathcal{A} &\equiv - T L_z  \mathcal{E}_q \sim 
(-i)\,\ln \det\,\big(i \partial\!\!\!/ - m - \beta H_{\rm int}\big)\,,
\label{Eq}
\end{align}
which is valid for static configurations in $H_{\rm int}$, is 
advantageous. The Feynman series generated via
\begin{align}
\mathcal{E}_{\rm FD}^{(n)}=\frac{\lambda^n}{n!}
\frac{i}{TL_z}\frac{\partial^n}{\partial \lambda^n}
\ln \det\,\big(i \partial\!\!\!/ - m - \lambda\beta H_{\rm int}\big)
\big|_{\lambda=0}
\nonumber\end{align}
is equivalent to the Born series; see appendix \ref{app:feyn} for more details. 
These Feynman diagrams are rendered finite when combined with 
standard counterterms whose contribution to vacuum polarization energy is 
$\mathcal{E}_{\rm CT}$. It remains to be observed that for the present model 
in $D=3+1$, the first $N=4$ Feynman diagrams are divergent. Hence $N=4$ Born 
subtractions are necessary to render the integral in eq.~(\ref{evac0}) finite:
\begin{align}
\mathcal{E}_q = - N_c \int_m^\infty \frac{dt}{4\pi}\,t \,\big[u(t)\big]_4 
+ \sum_{n=1}^4 \mathcal{E}_{\rm FD}^{(n)} + \mathcal{E}_{\rm CT}\,.
\label{evac1}
\end{align}
Here and in the following, the notation $[\ldots]_N$ indicates $N$ 
Born subtractions of scattering data inside the bracket. We stress that 
both the integral and the combined Feynman--counterterm contribution are 
individually finite. Thus no further (numerical) cut--off is required.

We have already mentioned that (in the numerical simulations) we measure
length scales in units of the inverse fermion mass $m$.  From 
Eqs.~(\ref{hfree},\ref{hint},\ref{evac0}) and~(\ref{Eq}) it then follows that
measuring the single particle energies and momenta in units of $m$ turns 
$\mathcal{E}_q$ into a dimensionless number that depends on any of 
the model parameters only via the counterterm
coefficients. Similarly the classical energy has a non trivial parameter 
dependence. Yet, the model parameters only enter local contributions to
the (total) energy, which are easy to compute. This simplifies
considerably the variational scan.

In principle, eq.~(\ref{evac1}) could be used directly to compute the vacuum
polarization energy. However, the exact calculation of the third- and fourth-order 
Feynman diagrams (including all finite parts) is very cumbersome. Fortunately, 
this is not really necessary: since the purpose of the Born subtraction is to render 
the spectral integral finite, we can subtract \emph{any} function with the correct 
asymptotic behavior, as long as we can associate this subtraction with a 
renormalizable Feynman diagram to be added back in. The third- and fourth-order 
Feynman diagram have a logarithmic divergence, which is also found for a 
second-order diagram of a simple scalar boson scattering off a background potential. 
If we adjust the size of this ``fake potential'' carefully, we can arrange for the 
logarithmic divergence in the second-order Boson diagram $\mathcal{E}_B^{(2)}$ 
to match the one from the fermion diagrams $\mathcal{E}_{\rm FD}^{(3)}+
\mathcal{E}_{\rm FD}^{(4)}$exactly. Instead of subtracting the third- and fourth-
order Born approximation and adding back in the corresponding fermion diagram, 
we can then subtract the (properly scaled) second Born approximation to 
a fake boson, and add back in the corresponding second-order boson diagram: 
\begin{align}
\mathcal{E}_q = - N_c \int_m^\infty \frac{dt}{4\pi}\,t\,\Big\{ 
2 \big[ u_F(t)\big]_2 + \frac{\lambda}{N_c}\, u_B^{(2)}(t) \Big\} 
+ \mathcal{E}^{(1,2)}_{F,\mathrm{ren}} + 
\lambda \mathcal{E}_B^{(2)} + \mathcal{E}_{\rm CT}^{(3,4)} \,.
\label{evac2}
\end{align}
Note the sign and the missing factor of 2 in the fake boson subtraction, 
which is due to the bosonic interface formula,\footnote{We have chosen the 
background potential in appendix \ref{app:fake} to be independent of $N_c$,
so the overall prefactor of $N_c$ is absent.}
\begin{align} 
\mathcal{E}_B^{(2)} = + \int_m^\infty \frac{dt}{4\pi}\,t\,u_B^{(2)}(t)\,.
\label{bosinter}
\end{align}
Next, we must choose the scaling factor $\lambda$ (not to be confused with
the Higgs coupling in Eq.~(\ref{model})) such that the logarithmic 
divergences in the fake boson and fermion diagram match:
\begin{align}
\lambda \equiv \frac{\big(\mathcal{E}_{\rm FD}^{(3)} + 
\mathcal{E}_{\rm FD}^{(4)}\big)\big\vert_\infty}{\mathcal{E}_B^{(2)}\vert_\infty}
= \frac{c_F}{c_B}
\end{align}
Here, $c_F$ and $c_B$ are simple radial integrals over the fermion profile 
functions or the fake boson potential, respectively, which parameterize the 
logarithmic divergence according to 
\begin{align}
\begin{rcases}
\begin{dcases}
\mathcal{E}_{\rm FD}^{(3)} + \mathcal{E}_{\rm FD}^{(4)} 
\\ \qquad \mathcal{E}_{\rm B}^{(2)}
\end{dcases}
\end{rcases} = 
- i  \pi \begin{rcases} \begin{dcases} c_F \\ c_B \end{dcases} \end{rcases}
m^{D-4} \int \frac{d^D k}{(2\pi)^D} \,\frac{1}{(k^2 - m^2 + i \epsilon)^2} 
+ \text{finite}\,.
\end{align}
Explicit formulae for $c_F$ and $c_B$ are listed in appendices
\ref{app:feyn} and \ref{app:fake}. Even without inspecting these
formulae, it is clear that $c_F$ is linear in $N_c$ because it originates
from a fermion loop. Hence $\frac{\lambda}{N_c}$ does not depend on 
$N_c$.

For the last step, we note that the fermion counterterms for the third- 
and fourth-order fermion diagram are, within the $\overline{MS}$ scheme, just 
the negative bare divergence 
$\mathcal{E}_{CT}^{(3,4)} = - \big(\mathcal{E}_{\rm FD}^{(3)} + 
\mathcal{E}_{\rm FD}^{(4)}\big)\big\vert_\infty$. Since this has been carefully matched 
to equal $-\lambda \mathcal{E}_B^{(2)}\big\vert_\infty$, the last two terms in 
Eq.~(\ref{evac2}) combine to the renormalized second-order fake boson diagram in $\overline{MS}$,
\begin{align}
\lambda \mathcal{E}_B^{(2)} + \mathcal{E}_{CT}^{(3,4)} = 
\lambda \Big[  \mathcal{E}_B^{(2)} - \mathcal{E}_B^{(2)}\big\vert_\infty\big]
= \lambda \,\mathcal{E}_B^{(2)}\big\vert_{\overline{MS}}\,.
\end{align}
Collecting all pieces, we can now rewrite the properly renormalized 
quantum correction to the energy per unit length of the string background as 
\begin{align}
\mathcal{E}_q &= - N_c \int_m^\infty \frac{dt}{4\pi}\,t\,\Big\{ 
2 \big[ u_F(t)\big]_2 + \frac{c_F}{c_B N_c}\, u_B^{(2)}(t) \Big\} 
+ \mathcal{E}^{(1,2)}\Big\vert_{\overline{MS}} + 
\frac{c_F}{c_B}\,\mathcal{E}_B^{(2)}\Big\vert_{\overline{MS}} 
+ \Delta \mathcal{E}_{\mathrm ren}
\nonumber\\[2mm]
&= \mathcal{E}_{\rm vac} + \big[\mathcal{E}_{\rm fermi}\big]_{\overline{MS}} +   
\big[\mathcal{E}_{\rm fake} \big]_{\overline{MS}}  + \Delta \mathcal{E}_{\rm ren}\,.
\label{evac3}
\end{align}
We note that simplifying the renormalization calculation by introducing the fake boson 
subtraction has been repeatedly tested for consistency. For example, in 
Ref.~\cite{Weigel:2016ncb} isospin and gauge symmetries were verified for 
$\mathcal{E}_q$ even though the individual terms on the right hand side  of
Eq.~(\ref{evac3}) are gauge variant. In Eq.~(\ref{evac3}) we have also added a finite 
counterterm contribution $\Delta \mathcal{E}_{\mathrm ren}$, which arises when we pass 
from the $\overline{MS}$ scheme to the more physical \emph{on-shell} scheme, such 
that the renormalized mass parameters agree with the actual physical particle masses. 
The contribution $\Delta \mathcal{E}_{\mathrm ren}$ contains the same terms 
as the classical energy Eq.~(\ref{Eclass}), but has different coefficients
computed from the finite parts of the second-order Feynman diagrams,
\begin{align} 
\Delta \mathcal{E}_{\rm ren} &= 
N_c\,\int_0^\infty dr\,r\,\Bigg\{\overline{c}_2\,\Big[s'(r)^2 + p'(r)^2 + 
\frac{p(r)^2}{r^2}\Big] + \overline{c}_4\,\Big[1 -s(r)^2 - p(r)^2\Big]^2\Bigg\}\,. 
\end{align}
Details on the coefficients $\overline{c}_2$ and $\overline{c}_4$ are presented 
in appendix \ref{app:onshell}. Eq.~(\ref{evac3}) is the master formula for the quantum 
energy of a neutral (uncharged) cosmic string. All four contributions are 
manifestly finite and well suited for numerical evaluation.

\subsection{Charged cosmic strings}
\label{sec:charge}
The quantum fluctuations computed from Eq.~(\ref{evac3}) usually do not lead to string 
stabilization. In fact, previous calculations \cite{Weigel:2009wi,Graham:2011fw} 
for Nielsen-Olesen type configurations showed that, at least for wide profiles,
the quantum corrections in $D=3+1$ have the \emph{same} sign as the classical 
energy. This implies that a stable string does not emerge, even when the quantum 
part is enhanced by {\it e.g.} assuming the heavy quark $f\to\infty$ or the large 
$N_c\to\infty$ limits.  Physically, this is not unexpected, as a negative total energy
would suggest that the vacuum is \emph{unstable} against cosmic string condensation. 

However, individual strings \emph{can} become bound if they manage to attract and bind 
sufficiently many fermions. In this scenario, fermions explicitly occupy bound states 
located near the string core, and the complete configuration is \emph{charged}, carrying 
the quantum number(s) of the trapped fermions. If the charge in question is conserved 
(at least to the extent that all charge-changing processes are suppressed by a large 
energy barrier), the charged string becomes (meta-)stable once its total energy is 
less than the masses of equivalently many free fermions.

More precisely, let $\epsilon_{i,\ell}$ be the eigenvalues of a square-integrable 
eigenstate of the single particle Hamiltonian, Eq.~(\ref{Ham}). Their computation 
is detailed in appendix \ref{app:charge}. Such bound states can occur in any angular 
momentum channel $\ell$. As the repulsion of the angular barrier increases with 
$\ell$, the number of bound states decreases and they disappear when $\ell$ is 
sufficiently large. We introduce a chemical potential $\mu$ and stipulate that all 
bound states $0 \le \epsilon_{i,\ell} \le \mu \le m$ are occupied explicitly. Emptying 
any of those levels and filling one that has $\epsilon_{i,\ell}>\mu$ only increases 
the energy. Assuming a quasi-continuum of states with energy $\sqrt{p^2 + \epsilon_{i,\ell}}$
and integrating over the momentum $p$ along the symmetry axis of the string
yields the charge per unit length \cite{Graham:2011fw}
\begin{align}
q(\mu) = \frac{N_c}{\pi}\sum_{0 \le \epsilon_{i,\ell} \le \mu} P_{i,\ell}(\mu)\,D_\ell\,,
\label{qmu}
\end{align}
where $P_{i,\ell}(\mu) = \sqrt{\mu^2 - \epsilon_{i,\ell}^2}$ is the Fermi momentum 
associated to a particular bound state of single particle Hamiltonian, Eq.~(\ref{Ham}). We
have also included the degeneracy $N_c D_\ell$ of each state due to angular momentum 
and color. As discussed after Eq.~(\ref{evac1}), the charge per unit length is measured in 
multiples of $m$, as are the bound state energies and the chemical potential. Next we invert 
the relation in Eq.~(\ref{qmu}) to compute $\mu_Q$, for a prescribed charge per unit
length and calculate the binding energy per unit length is calculated. In practice this
requires three steps: 
\begin{enumerate}
 \item prescribe a value $Q \ge 0$ for the charge per unit length;
 \item determine the chemical potential $\mu_Q$ by increasing $\mu$ in small steps, 
  starting at $\mu = \min_{i,\ell} |\epsilon_{i,\ell}|$, until the 
  condition $q(\mu_Q) = Q$ is met
  or $\mu > m$ (whence the chosen charge $Q$ cannot be accommodated);
 \item  Then, sum over all single particle bound states, integrate over $p$
 up to the Fermi momentum and subtract $q(\mu_Q)m$, the equivalent energy of 
 free fermions, to obtain the binding energy per unit length \cite{Graham:2011fw}
 \begin{align}
 \mathcal{E}_b(Q) &= N_c \sum_{0 \le \epsilon_i \le \mu} \int_0^{P_i(\mu)} 
 \frac{dp_z}{\pi} \,\Bigg[ \sqrt{\epsilon_i^2 + p_z^2} - m\Bigg]\,D(\epsilon_i) 
 \nonumber \\[2mm]
& = \frac{N_c}{2\pi}\sum_{0 \le \epsilon_{i,\ell} \le \mu_Q} 
 \Bigg[ P_{i,\ell}(\mu_Q)\,(\mu_Q - 2m) + \epsilon_i^2\,\ln \frac{P_{i,\ell}(\mu_Q) + \mu_Q}
 {\epsilon_{i,\ell}}\Bigg]\,D_\ell\,. 
 \label{Eb}
 \end{align}
\end{enumerate}

Since $\mathcal{E}_b(Q) < 0$ by construction, charging the string always has a 
binding effect, though it may not be strong enough to overcome the other 
contributions to the total energy.  In addition, the total number of bound states 
in a given string background is finite, so that there is a maximal charge per 
unit length $Q_{\rm max} = q(m)$ that can be placed on the string,
and hence also a limit  to the binding effect generated by charging the string.

Equations (\ref{evac3}) and (\ref{Eb}) comprise all contributions to the 
quantum energy of a hedgehog type of cosmic string, at least in the limit 
$N_c \to \infty$ when the fermion determinant dominates all quantum corrections.
Since $\mathcal{E}_q$ and $\mathcal{E}_b(Q)$ saturate the $\mathcal{O}(N_c\hbar)$ 
contribution to the energy, any consideration of $\mathcal{E}_b(Q)$ requires the 
inclusion of $\mathcal{E}_q$ for consistency.

\section{Numerical studies and results}
\label{sec:results}
In this chapter we present the numerical results of our investigation. In the first 
part, we discuss the individual contributions to the string energy separately, and 
perform numerical tests on their computation. In the second part we report the 
results of our variational search for optimal string profile  parameters. In all 
calculations we employ the hedgehog ansatz, Eq.~(\ref{config1}) consistent with 
the boundary conditions derived in Eq.~(\ref{103}). We introduce two variational 
width parameters $w_r$ and $w_a$ for the background profiles
\begin{align}
 \rho(r) = 1 - a \cdot \exp\left(-\frac{r^2}{2 w_r^2}\right)\,,\qquad\qquad
 \theta(r) = - \pi\cdot \exp\left(-\frac{r}{w_a}\right)\,,
 \label{ansatz}
\end{align}
of the chiral radius and chiral angle, respectively. The amplitude $a$ describes 
the decrease in the Higgs condensate at the core of the string: $\rho_0=1-a$.
Inspired by the Nielsen-Olesen profiles this amplitude is often chosen as $a=1$ so 
that $\rho_0=0$. This results in strongly bound states, since fermions located 
in the vicinity of the string core have near zero mass. 
Taking $a\to1$ produces more ``shallow'' profiles. Though they produce 
less deeply bound states, a non-zero $a$ may nevertheless be beneficial in reducing 
the total energy because its smaller gradients decrease the classical energy.
The complete ansatz, Eq.~(\ref{ansatz}) thus comprises three variational
parameters $a$, $w_r$ and $w_a$.

\subsection{Numerical details for a single string background}

\renewcommand{\arraystretch}{1.0} 
\begin{table}[h!]
	\centering
	\begin{tabular}{ccccc}
		\toprule
	    contribution & comment & equation  & depends on & value [$m^{-2}$]
	    \\
	    \colrule
	    $\mathcal{E}_{\rm cl}$ &  
	    \parbox{6cm}{classical energy} & 
		(\ref{Eclass}) & \parbox{4cm}{Yukawa coupling $f$ } & $14.96$
		\\
	    $\mathcal{E}_{\rm FD}^{(1,2)}\vert_{\overline{MS}}$ &  
        \parbox{6cm}{$2^{\rm nd}$ order fermion diagram} & 
        (\ref{app_Efermi}) & --- & $-0.13$
        \\
	    $\Delta\mathcal{E}_{\rm ren}$ &  
	    \parbox{6cm}{finite counterterm $\overline{MS} \to \rm{on shell}$} & 
        (\ref{app_Eren}) & Yukawa coupling $f$ & $0.27$
        \\
	    $\mathcal{E}_{\rm B}^{(2)}\vert_{\overline{MS}}$ &  
        \parbox{6cm}{$2^{\rm nd}$ order fake boson diagram} & 
        (\ref{app_Efake}) & --- & $0.02$
        \\
	    $\mathcal{E}_{\rm vac}$ &  
	    \parbox{6cm}{vacuum polarization energy} & 
        (\ref{evac3}) & --- & $0.94$
        \\
	    $\mathcal{E}_{b}$ &  
	    \parbox{6cm}{charge energy} & 
        (\ref{Eb}) &string charge $Q$ & $-1.85$
        \\ \colrule
	    $\mathcal{E}_{tot}$ &  
	    \parbox{6cm}{total energy per unit length} & 
          & $(f,Q)$ & $14.21$
		\\ \botrule
		\vspace{-6mm}
	\end{tabular}
	\caption{\label{tab:1}Contributions to the total energy per unit length for a sample 
	hedgehog string background with parameters $w_a = w_r = 3/m$ and $a=1$.
    The model parameters are taken from the physical top quark and Higgs masses: $f=0.99$ and 
    $\lambda = 0.129$, {\it cf.} Eqs.~(\ref{mtop}) and~(\ref{mHiggs}). The
    charge per unit length is set to a typical value of $Q=5m$.}
\end{table}
\renewcommand{\arraystretch}{1.0}

In this section, we survey our numerical procedure for a single background configuration 
for the case of
\begin{align}
w_r = w_a = 3/m\qquad{\rm and}\qquad a = 1\,.
\label{para1}
\end{align}

The total energy per unit length of the string background in our framework comprises six 
contributions, which are listed in table \ref{tab:1}. For a fixed string background, only the 
classical energy and the finite counterterm (which is always significantly smaller than the 
classical part) depend on the Yukawa coupling $f$, since the fermion determinant contribution,
$\ln \det\,\big(i \partial\!\!\!/ - m - \beta H_{\rm int}\big)$, is independent 
of $f$ when all energies are measured in units of $m$. Increasing the Yukawa coupling 
$f$, {\it i.e.} the ratio between the fermion mass and the Higgs \emph{vev}, reduces the
classical contribution so if the net contribution of the  quantum corrections is negative, we 
can always get a stable string by increasing $f$ to the point where the classical energy penalty 
becomes negligible. From table \ref{tab:1}, we recognize that this mechanism requires 
having the string carry charge, since the remaining quantum corrections ({\it i.e.}~the pure 
fermion determinant) is typically positive and hence does not cause binding. The energy from 
Eq.~(\ref{evac3}) may indeed become negative for large Yukawa coupling and very narrow profiles 
($w_r,w_a\to0$) \cite{Graham:2011fw}. The Fourier momentum of such profiles then approaches the 
Landau ghost \cite{Ripka:1987ne}, indicating that the one loop approximation fails. We thus ignore 
configurations that are afflicted by the Landau ghost problem.

Of all the contributions shown in table \ref{tab:1}, only the vacuum polarization and 
the charge energy are numerically expensive to compute. The remaining pieces are just simple integrals 
in coordinate or momentum space. We will now present some numerical details on the computation 
of these expensive contributions:

\begin{figure}[t]
	\centering
	\includegraphics[width=7cm,clip]{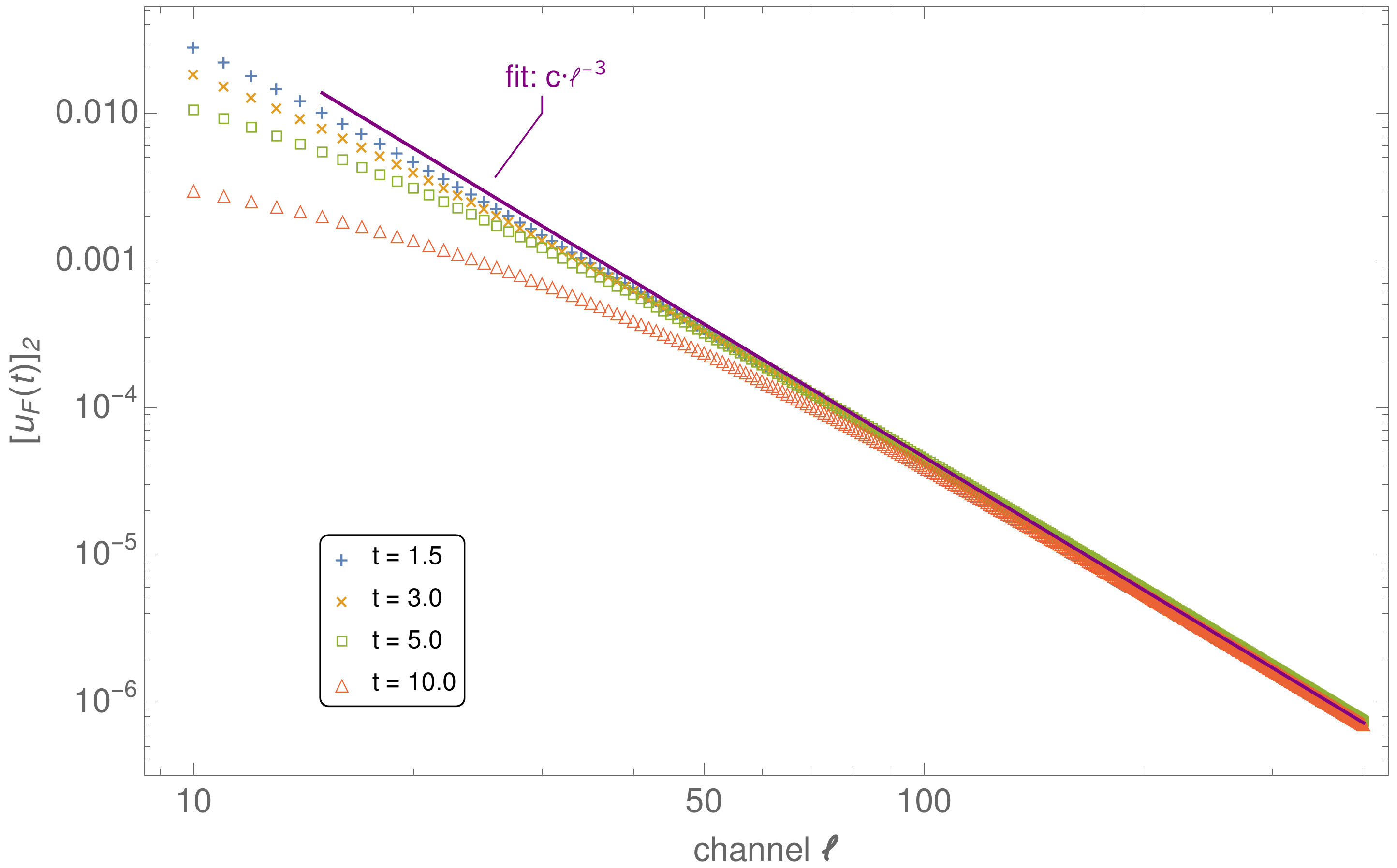}
	\hspace*{1cm}
	\includegraphics[width=7cm,clip]{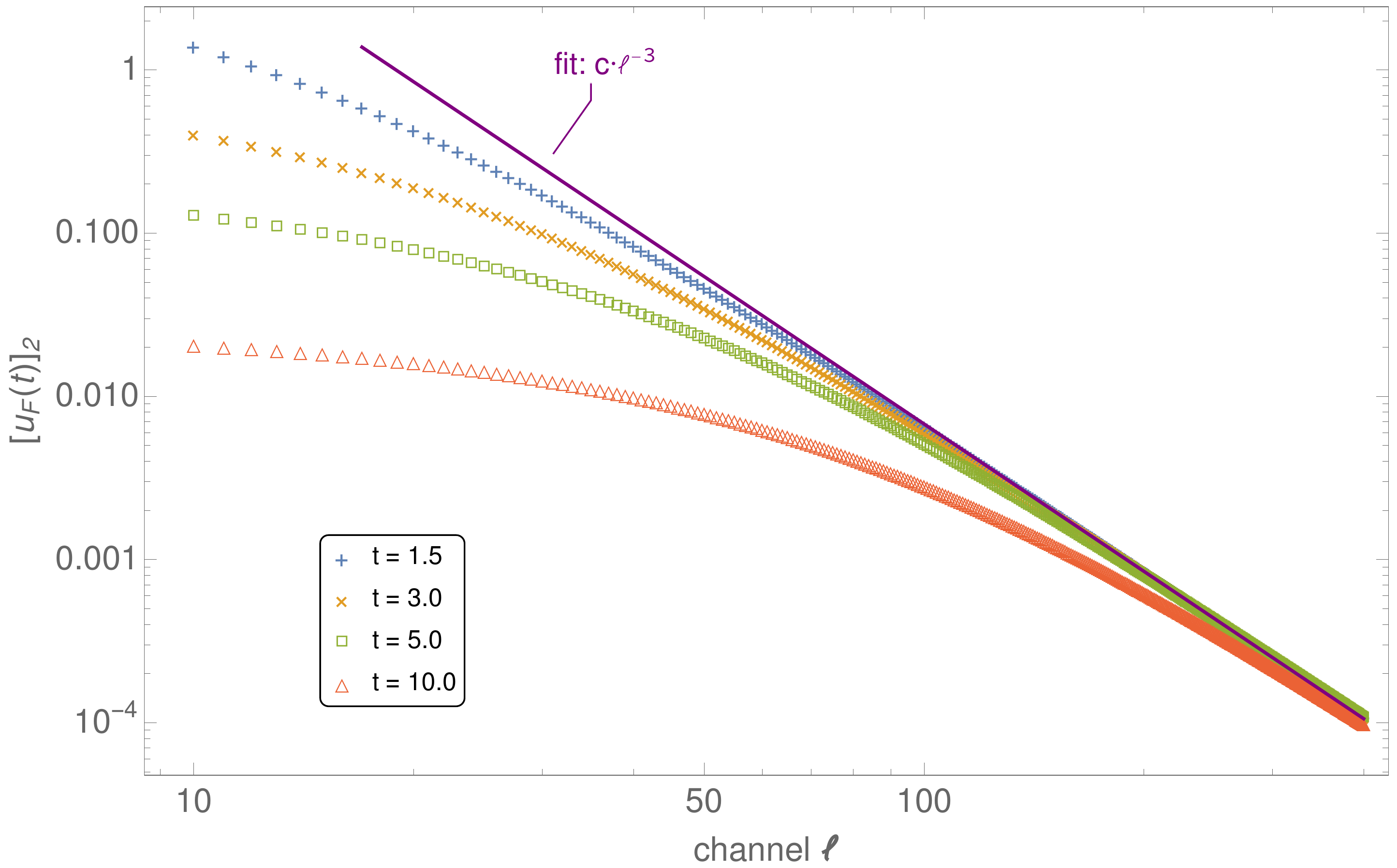}
	\caption{The contributions to the twice Born subtracted channel sum $u_F(t)$ in 
		eq.~(\ref{defu}), for various imaginary momenta $t$. The left chart corresponds
	to a narrow string with $w_a = w_r = 2/m$, while the right chart shows 
    the case of a wide string with $w_a = w_r = 7/m$. As can be seen, wider strings 
    generally require more channels to reach the 
    asymptotic region with a power law decay. Also, the shift 
    of the asymptotic region to larger channels with increasing momentum is 
    much more pronounced for wider strings.}
	\label{fig:1}       
\end{figure}

\subsubsection*{Vacuum energy}

The main ingredient for the vacuum polarization energy $\mathcal{E}_{\rm vac}$ in
Eq.~(\ref{evac3}) is the sum over the twice Born subtracted logarithm of the Jost 
function, $D_\ell \,[\nu_\ell(t)]_2$, defined in Eq.~(\ref{defu}). Its numerical 
evaluation is costly because many angular momenta must be included. We present 
a double logarithmic plot of $D_\ell \,[\nu_\ell(t)]_2$ in Fig.~\ref{fig:1}.  
As can be seen, the individual contributions eventually 
decay with power law $\ell^{-3}$, which allows for the use of series accelerators.
Still, at least $200$ channels, and up to $500$ channels at higher momenta, 
need to be summed to get an accurate estimate of $\left[u_F(t)\right]_2$, and 
likewise for the fake boson contribution $u^{(2)}_B(t)$.

In order to further analyze the vacuum energy, we separate the integrand in 
$\mathcal{E}_{\rm vac}$ into the fermion and fake boson parts,
\begin{align}
s_F(\tau) &\equiv - \frac{N_c}{4 \pi}\,\tau \,\left[2u_F(\sqrt{\tau^2+m^2})\right]_2\,,
\nonumber \\[2mm]
s_B(\tau) &\equiv  + \frac{N_c}{4 \pi}\,\tau \frac{c_F}{c_B N_c}\,u_B^{(2)}(\sqrt{\tau^2+m^2})\,,
\label{defs}
\end{align}
where we have also changed the momentum variable $t \to \tau \equiv \sqrt{t^2 - m^2}$.
Here, each function $u(t)$ is the sum of the logarithmic Jost function over 
all angular momenta, {\it cf}.~Eqs.~(\ref{defu}) and~(\ref{defuB}). 

\begin{figure}[t]
	\centering
	\includegraphics[width=7cm,clip]{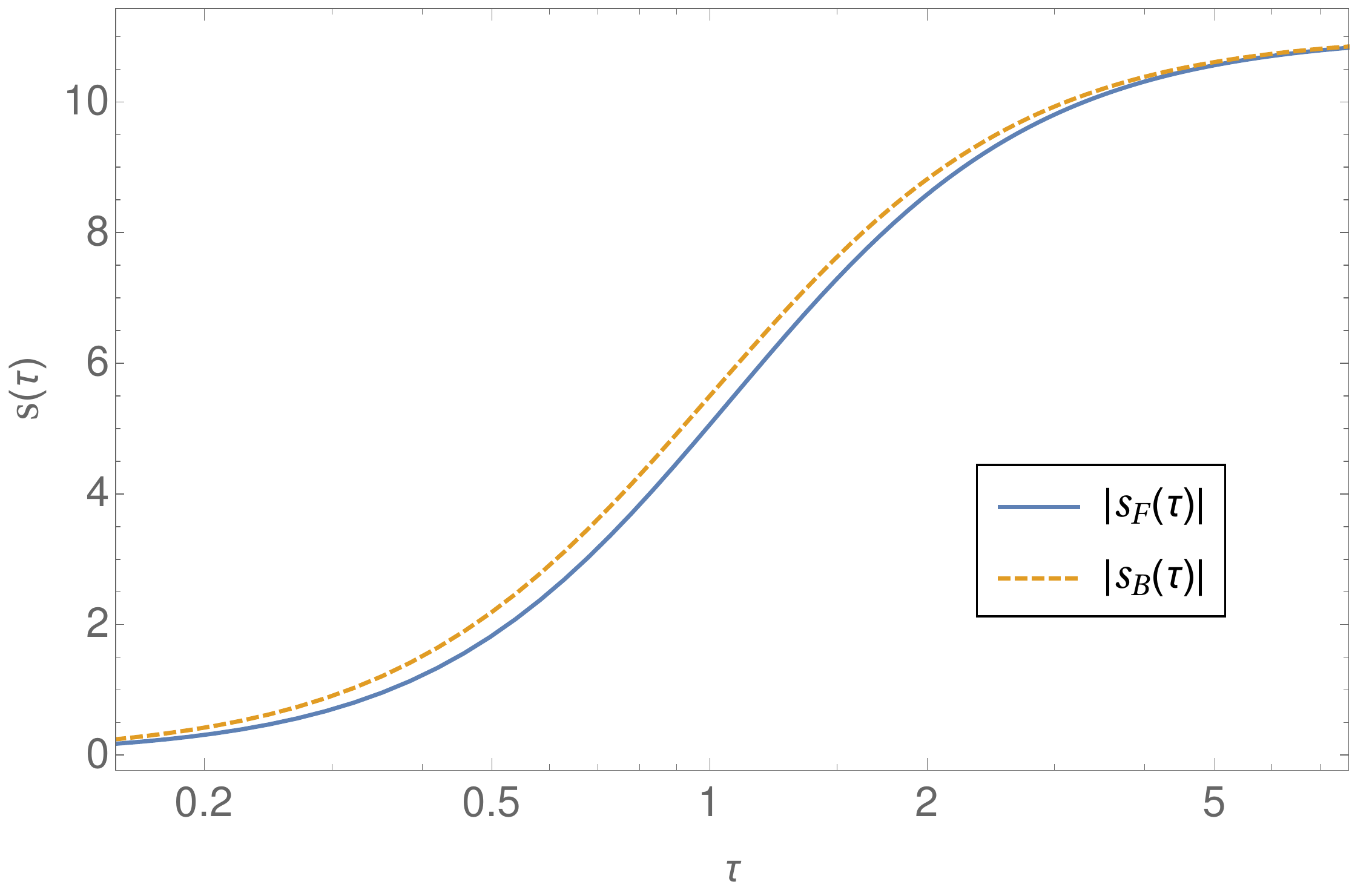}
	\hspace*{1cm}
	\includegraphics[width=7cm,clip]{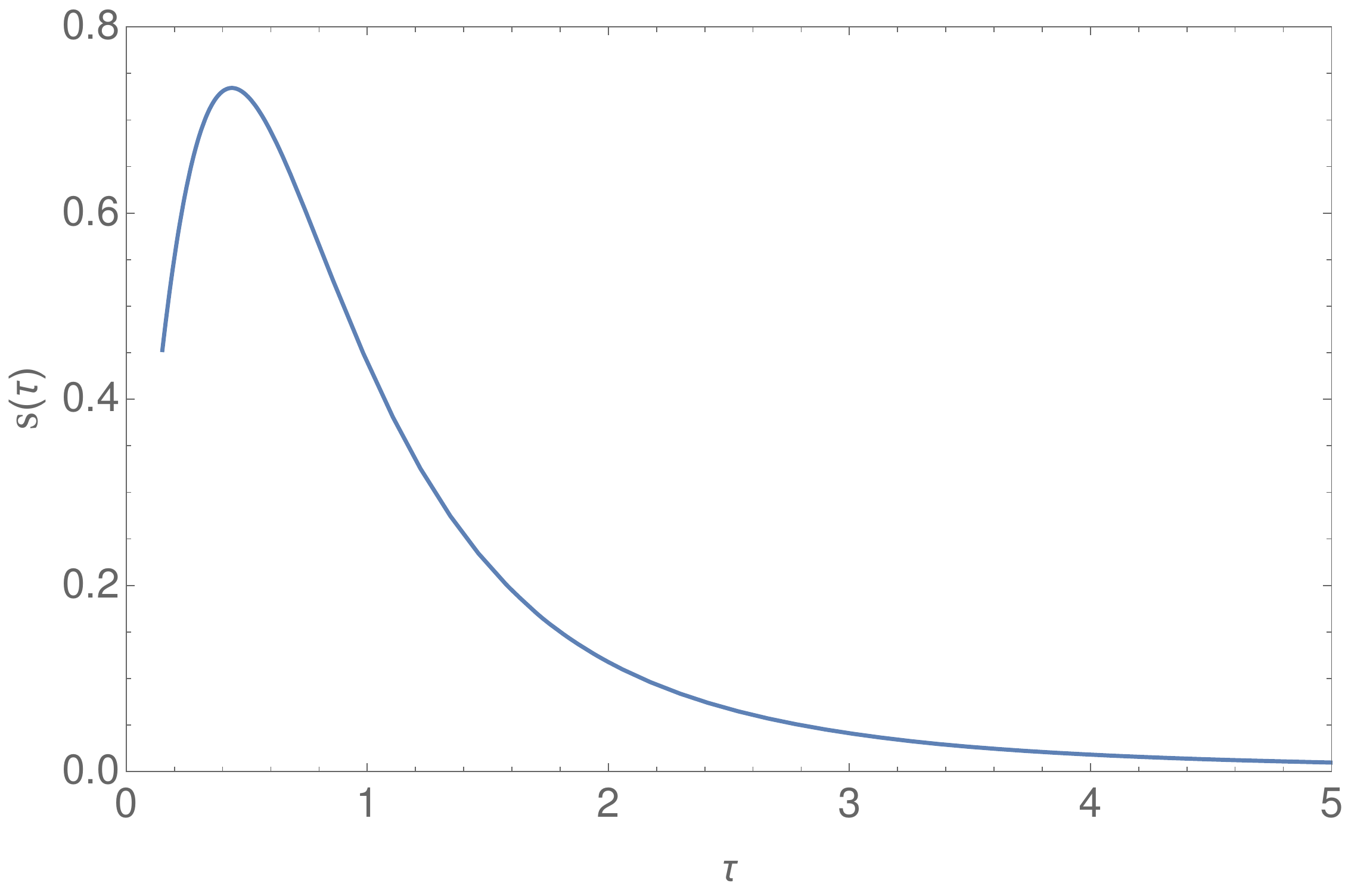}
	\caption{\emph{Left}: The fermionic and fake boson contributions to the 
		integrand eq.~(\ref{defs}) for the vacuum polarization energy. 
		For clarity, the contribution is multiplied by $\tau$ to emphasize 
		the asymptotic decay  $\sim 1/\tau$. \emph{Right:} The full
		integrand $s(\tau)$ eq.~(\ref{defs}) for the vacuum energy.}
	\label{fig:2}       
\end{figure}

The fake boson method relies on the fact that a properly rescaled second-order boson 
contribution possesses the same logarithmic divergence as the third- and fourth-order 
Feynman diagrams, {\it i.e.} the large-momenta behavior of the two integrands 
$s_F(\tau)$ and $s_B(\tau )$ in Eq.~(\ref{defs}) must match. In the left panel of 
Fig.~\ref{fig:2}, we present the products $\tau \,s_F(\tau)$ and $\tau\,s_B(\tau)$, 
because they should asymptotically approach the (same) constant in order to cancel the 
(same) logarithmic divergence in $\mathcal{E}_{\rm vac}$. This is indeed the case to a 
very high accuracy. Though the full calculation is computationally expensive, it has the 
advantage to  provide an independent test for the precision of our numerics.

In the right panel of Fig.~\ref{fig:2}, we show the complete integrand 
$s(\tau)  \equiv s_F(\tau) - s_B(\tau)$ of the integral in $\mathcal{E}_{\rm vac}$.
With the fake boson subtraction, the integrand vanishes very quickly already for 
moderate momenta, which allows for an accurate evaluation\footnote{We truncate the 
$\tau$-integral at a very small and a very large cutoffs and estimate the remainder 
in both regions by fits to the integrand which are then extrapolated and integrated
analytically. At small $\tau$, we use a quadratic polynomial fit, while at large 
momenta $\tau \gg 1$, we assume a power-law decay. The cutoffs are determined
such that the low- and high-momentum extrapolations are less than $5\%$ of the bulk 
contribution. Stability of this procedure against moderate variations of the cutoffs 
is verified.} of $\mathcal{E}_{\rm vac} \approx 0.94$, as listed in table~\ref{tab:1}.

\subsubsection*{Bound state energy}

We compute matrix elements of the Hamiltonian $H=H_0+H_{\rm int}$,
Eqs.~(\ref{hfree}) and~(\ref{hint}) with respect to the eigenfunctions of
$H_0$. The details of this calculation are described in appendix \ref{app:charge}.
The would-be scattering and shallow bound states near threshold will still vary 
considerably with the artificial numerical parameters; but the real bound-state spectrum 
of eigenvalues $<0.95m$ is stable.  In table \ref{tab:2}, we list the positive bound 
states for all angular momentum channels for the string background with 
$w_a=w_r=3/m$. As discussed above, the interaction is charge conjugation 
invariant, so for each positive energy solution there is a negative one.

\renewcommand{\arraystretch}{1.0} 
\begin{table}[t]
	\centering
	\begin{tabular}{ccl}
		\toprule
		\parbox{4cm}{channel index $\ell$} & 
	    \parbox{4cm}{\# bound states} & 
        \parbox{5cm}{positive bound state energies}
		\\ \colrule 
		$-1$ & 14 & $0.133, \quad 0.601,\quad 0.702,\quad 0.807, \quad 0.903, \quad 0.930, \quad 0.970 $
		\\
		$0$  & 10 & $0.427,\quad 0.616,\quad 0.807,\quad 0.866,\quad 0.996$
		\\
		$1$  &  8 & $0.672, \quad 0.859, \quad 0.957,\quad 0.973$
		\\
		$2$  &  2 & $0.862$
		\\
		$>2$ &  0 & ---
		\\ \botrule
	\end{tabular}
        \caption{\label{tab:2}Energies and angular momenta of the 
        fermion bound states in the background of the hedgehog soliton with $w_a = w_r = 3/m$ and $a=1$.}
\end{table}
\renewcommand{\arraystretch}{1.0}

With the bound states at hand, we can evaluate the binding effect from charging the string as 
laid out in section \ref{sec:charge}. Here we first report the the maximal charge (per unit 
length) which the string with the parameters from Eq.~(\ref{para1}) can accommodate. It is 
obtained by equating the chemical potential with the fermion mass in Eq.~(\ref{qmu}): $q(m)=13.78$.

\begin{figure}[t]
	\centerline{
	\includegraphics[width=7cm,clip]{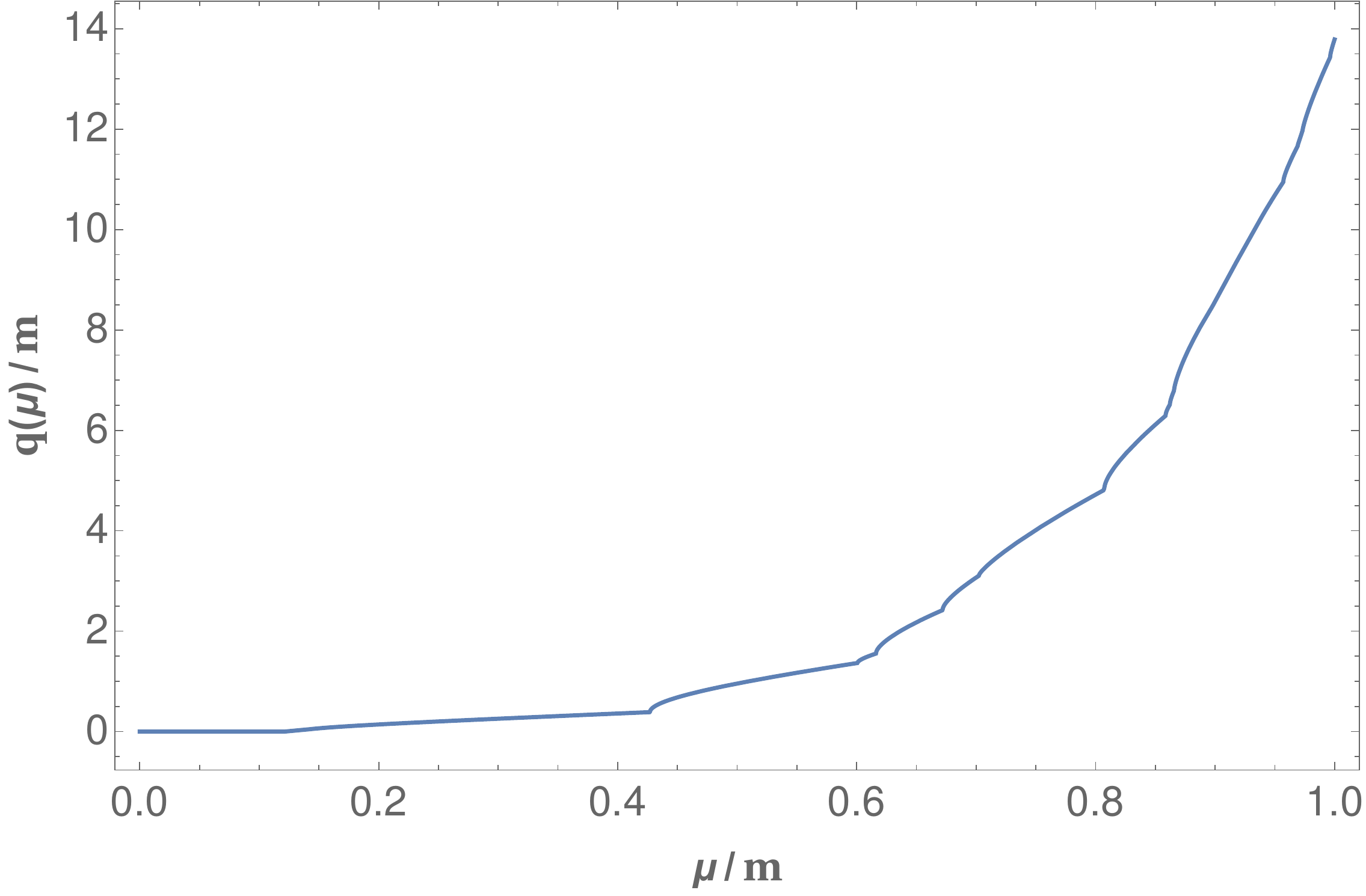}
	\hspace*{1cm}
	\includegraphics[width=7cm,clip]{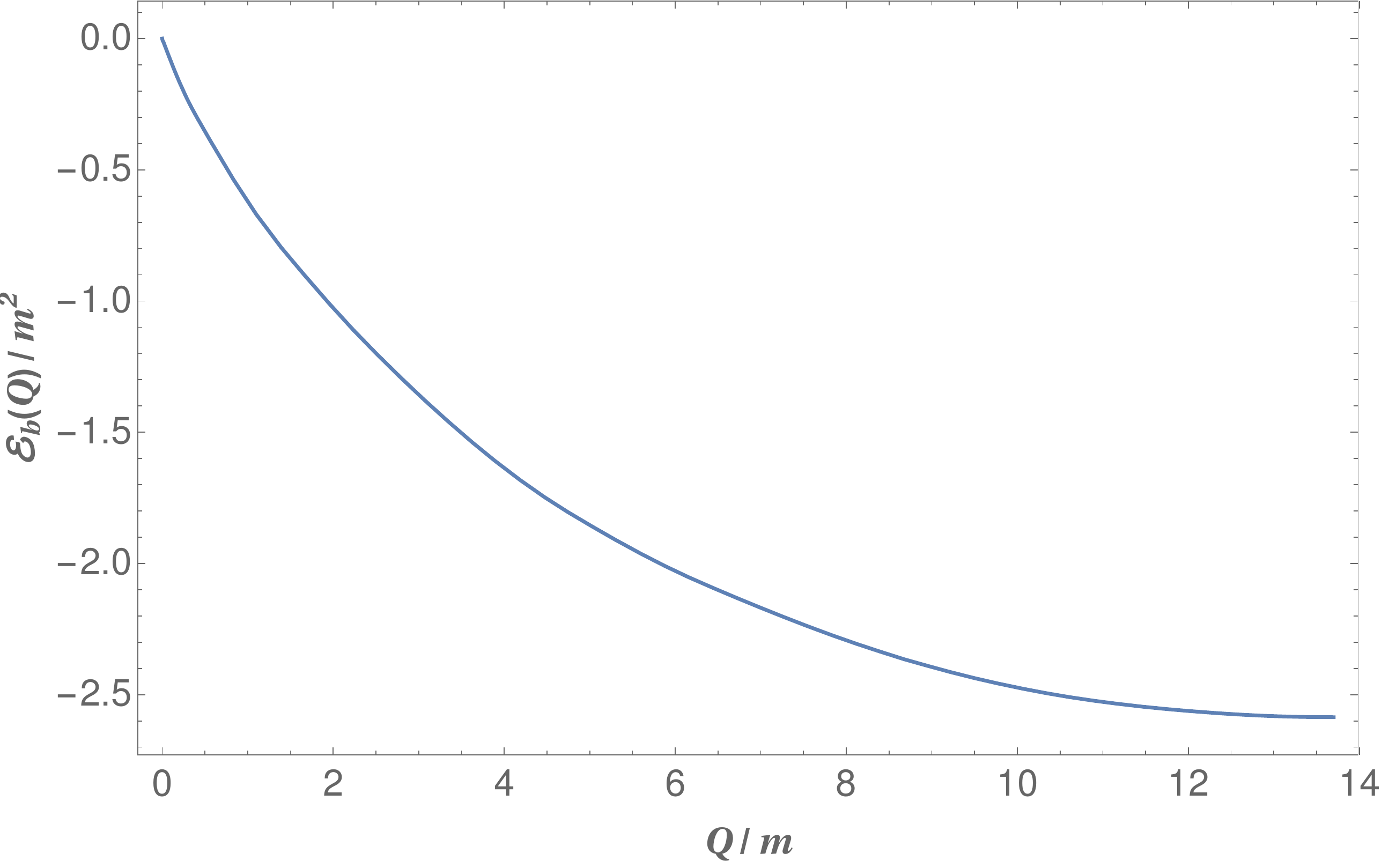}
        }
	\caption{\emph{Left}: The charge per unit length $q(\mu)$ induced on the 
		hedgehog string by filling all levels lower than a chemical 
		potential $\mu$, for the string background with $w_a = w_r = 3/m$.
		\emph{Right}: The binding energy $\mathcal{E}_b(Q)$
        due to a prescribed charge per unit length $Q$, for the same
		string background.}
	\label{fig:3}       
\end{figure}

Secondly, we plot the charge, Eq.~(\ref{qmu}), as as function of the chemical potential in 
Fig.~\ref{fig:3}. It is monotonically increasing by construction and can be inverted numerically 
to yield the chemical potential $\mu(Q)$ necessary to produce a given charge $Q$. With this relation, 
the energy per unit length $\mathcal{E}_b(Q)$ induced by the string charge $Q$ can be evaluated 
from Eq.~(\ref{Eb}). For our model string, Eq.~(\ref{para1}), $\mathcal{E}_b(Q)$ is plotted in the 
right panel of Fig.~\ref{fig:3}. By construction, $\mathcal{E}_b(Q)$ is negative and monotonically 
decreasing up to the maximal charge $q(m)$ allowed by the Pauli principle. Fig.~\ref{fig:3} shows 
that the binding energy due to a maximally charged string is $\mathcal{E}_b(q(m))= -2.5\,m^2$, 
while a realistic value for a moderate charge $Q\approx 5m$ is $\mathcal{E}_b(Q)=-1.85\, m^2$.

\subsubsection*{Total energy}

\begin{figure}[t]
	\centering
	\includegraphics[width=12cm,height=5cm,clip]{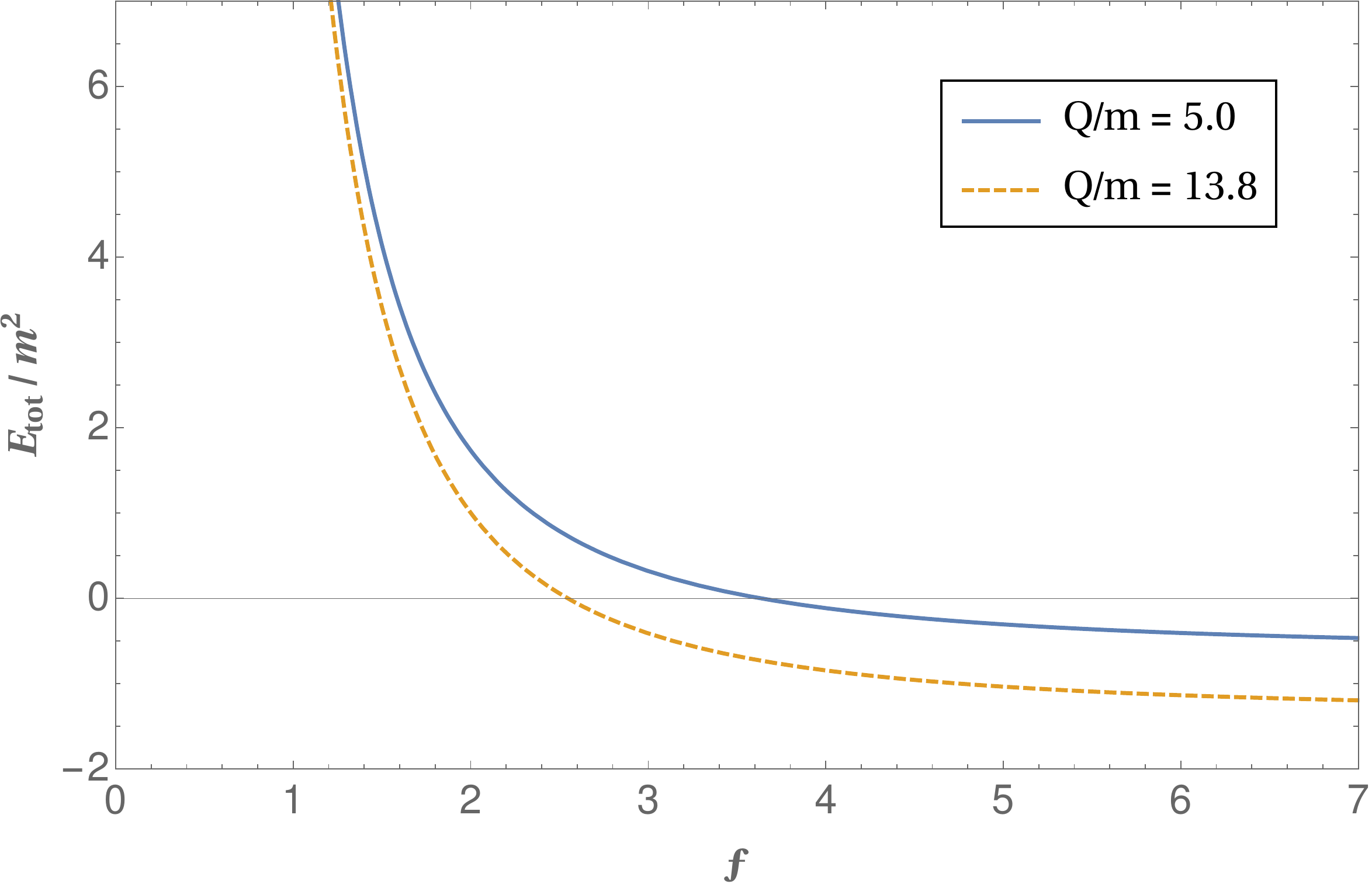}
	\caption{The total energy per unit length, Eq.~(\ref{etot}) of a cosmic string background
		with widths $w_a = w_r = 3/m$ as a function of the Yukawa coupling
		at fixed charge per unit length.}
	\label{fig:4}       
\end{figure}

Comparing the binding effect of charging the string with the remaining contributions 
to the string energy in table \ref{tab:1}, obviously shows that the charged string with 
$w_a = w_r = 3/m$ is \emph{not} stable when the Yukawa coupling $f$ is adjusted to
the physical top quark mass. A slight increase of $f$ to reduce the large classical 
energy as in Fig.~\ref{fig:1} indeed gives a bound object. In Fig.~\ref{fig:4}, we show 
the total energy per unit length 
\begin{equation}
E_{\rm tot}(Q)=\mathcal{E}_{\rm cl}+\mathcal{E}_{q}+\mathcal{E}_b(Q)
\label{etot}
\end{equation}
of a charged string  with variational parameters $w_a=w_r=3/m$ as a function of the 
Yukawa coupling $f$. For a moderate charge, $Q=5\,m$, the string becomes bound around 
$f\approx 3.66$, which corresponds to a fermion mass of $m = 637\,\mathrm{GeV}$ (assuming 
the empirical \emph{vev}, $v=174\,{\rm GeV}$). If instead we allow the string to be maximally 
charged, the threshold for binding drops to $f\approx 2.55$ corresponding to a fermion mass 
of $m=443\,\mathrm{GeV}$. 

\subsection{Variational searches for bound cosmic strings}

\begin{figure}[t]
	\centering
	\includegraphics[width=12cm,height=5cm,clip]{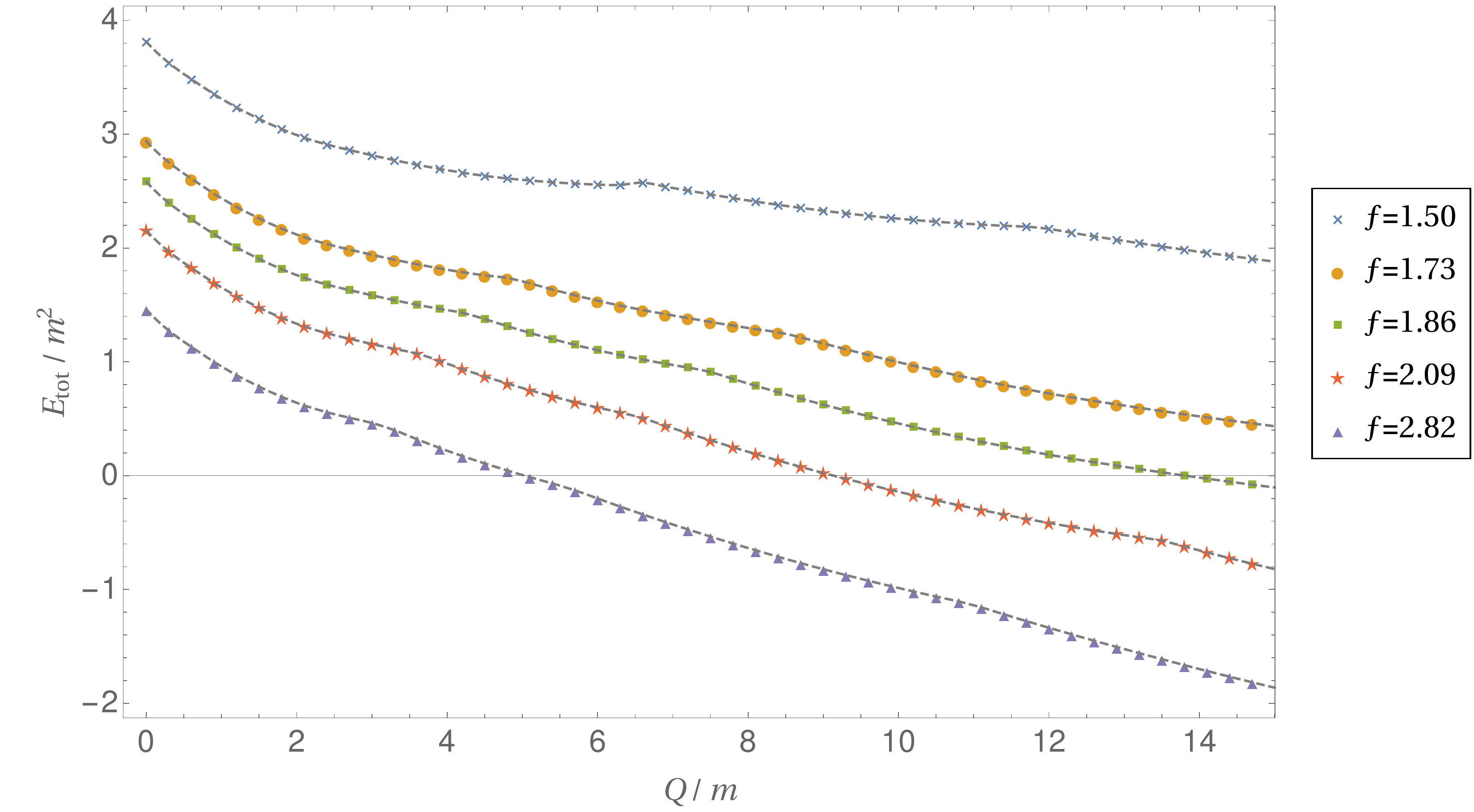}
	\caption{The total energy per unit length of the optimal string configuration, 
		as a function of the string charge per unit length, $Q$, and for various 
		values of the Yukawa coupling $f$.}
	\label{fig:5}       
\end{figure}

The results for the single configuration presented above are representative for a 
typical string background. They give  an upper limit on the fermion mass needed
to bind a (charged) cosmic string. We can improve this limit by varying the 
variational parameters of the background profile to identify the optimal string 
shape for any given coupling or charge. For this purpose, we have varied the width 
parameters $w_a$ and~$w_r$ in the string profile eq.~(\ref{ansatz}) within the 
range $w_a, w_r \in [1/m,10/m]$. Smaller values may yield a lower
$\mathcal{E}_b$ as an artifact of the Landau pole and are therefore
discarded. In addition to testing (several hundred) configurations that all have 
a vanishing Higgs background at the string core, we have also included about 30 
``shallow'' configurations with amplitude parameter $a \in [0.1, \,0.9]$ 
in the set of sample string profiles. For each of these 
configurations we have computed the vacuum polarization energy and the bound state 
spectrum.  We then select a value for the Yukawa coupling $f$ and compute the total 
binding energy $E_{\rm tot}(Q)$, Eq.~(\ref{etot}) as a function of $Q$, the string 
charge per unit length, for all configurations. Finally, at any given $Q$ we determine 
the minimal $E_{\rm tot}(Q)$. In Fig.~\ref{fig:5}, we show the final result of this 
variational search.  Typically a particular configuration is optimal for a finite 
interval in $Q$. When $Q$ is increased eventually the maximal charge $q(m)$ that this 
configuration can accommodate is reached and a switch occurs to another optimal 
configuration that can hold a larger charge.  This switching of optimal configurations 
gives rise to small bends in the curves. For small Yukawa couplings, the total binding 
energy stays positive and no stable string is found. As we increase the Yukawa coupling, 
the total binding energy decreases for large $Q$ and eventually turns negative. We find 
that the smallest Yukawa coupling, for which a stable charged string is observed
is $f\approx1.86$. This corresponds to a quark mass of $m\approx324\,\mathrm{GeV}$. 
This binding occurs at an almost maximal charge per unit length of $Q\approx13\,m$. As we 
further increase the Yukawa coupling, less and less charge is necessary to obtain a bound 
string. At $f \approx 2.82$ or a quark mass of $m \approx 490\,\mathrm{GeV}$, a relatively 
moderate charge of $Q \approx 5\,m$ is sufficient to bind the cosmic string.

\begin{figure}[t]
	\centering
	\includegraphics[width=10cm,height=8cm,clip]{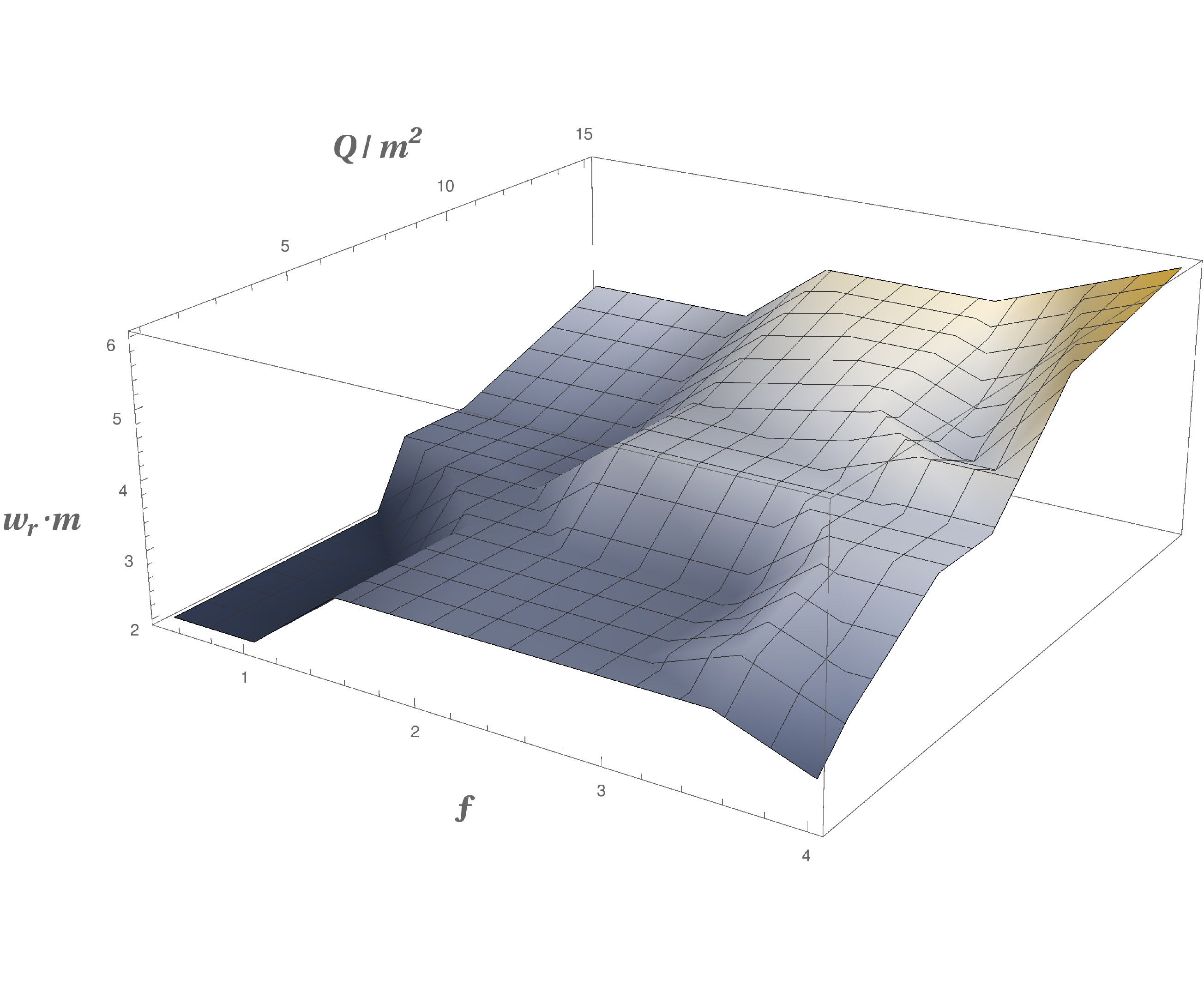}
	\caption{The radial width $w_r$ of the optimal string configuration (in units of 
	         $m^{-1}$) for various charges and Yukawa couplings at a fixed 
	         angular width $w_a=2/m$.}
	\label{fig:4c}       
\end{figure}

\medskip\noindent
We find four general features of the optimal string configuration:
\begin{enumerate}
\item All optimal configurations have $a=1$, {\it i.e.}~it is preferable to have the 
Higgs field vanish at the origin, as in the Nielsen-Olesen profile. This is somewhat
unexpected as it contrasts with the motivation for the hedgehog configuration, Eq.~(\ref{config}).
The profiles with $a=1$ have fewer, but deeper bound states and a considerable classical 
energy. The ``shallow'' configurations with a non-vanishing Higgs condensate at the string 
core are not optimal, even though they cost less classical energy to form. Since for 
shallow configurations all bound states are close to threshold, the loss in binding 
energy at large charges outweighs the gain in classical energy. 
\item  All optimal configurations have $w_a = 2$, {\it i.e.}~the angular 
twisting of the Higgs emerges close to the string core, even when the radial distribution
of the string profile is rather wide.\footnote{We have also investigated configurations 
with smaller $w_a = 1.1$ and $w_a = 1.5$, which were not optimal, so that the value 
$w_a = 2$ is not a corner case.}
\item The width of the radial Higgs profile generally increases with increasing charge $Q$,
as can be seen from Fig.~\ref{fig:4c}. Since wider strings bind 
charge more easily, 
the optimal configurations are fairly wide for the lightest possible quarks masses. However, 
we have included radial widths up to $w_r = 10/m$ in our variational search, and extremely 
wide configurations with $w_r \ge 7/m$ are not preferable.  
\item For $f > 1.86$, we find bound strings at a critical charge $Q > Q^\ast$, which 
\emph{decreases} with increasing quark mass. At the same time, the radial width of the chiral 
radius of the optimal configuration for the critical charge $Q^\ast$ actually decreases for 
higher fermion masses, {\it e.g.} from $w_r^\ast=4.0/m$ at $f=1.86$ to
$w_r^\ast=1.90/m$ at $f=5.0$.   
\end{enumerate}

\section{Summary and conclusions}
\label{sec:summary}

We have investigated the dynamical stabilization of a cosmic string in an $SU(2)$ gauge 
theory that is a slightly reduced version of the electroweak standard model. The string 
configuration itself consists of a twisted string-like deviation from the Higgs \emph{vev} 
without any gauge field admixture, {\it i.e.}~a thin line defect carved into the Higgs condensate. 
This ansatz is inspired by the well-known hedgehog ansatz for the chiral soliton in 
quark models. In contrast to the Nielsen-Olesen configuration, the present one is 
characterized by two profile functions for the Higgs field, a chiral radius and a chiral 
angle. The latter is similar to the Skyrme model solution.  Classically, the string 
configuration is not stable, but it tends to attract fermions which may be bound in the 
vicinity of the string core to produce a \emph{charged} string. As a consequence the charged 
string becomes stable if the quark mass is large enough. For consistency of the $\hbar$ 
expansion we must also include the contribution of the scattering states to the quantum 
energy, and renormalize conventionally to make contact with empirical model parameters. 
This is the most complicated and numerically expensive part of the calculation.

We find that at a fairly large charge the string becomes bound when the fermion mass 
exceeds a value of about $320\,\mathrm{GeV}$. By charge conservation it can only decay 
into a system of equally many free fermions which, however, has a bigger energy. The 
resulting string profiles are characterized by a fairly narrow chiral angle that has a 
width of about $w_a = 2/m$ while the chiral radius is more extended with a width $w_r=4/m$. 
To put this in perspective, consider the optimal string at the smallest possible fermion 
mass of $320\,\mathrm{GeV}$. If it extends over a length equal to the diameter of the sun, 
the mass of the optimal string would only be a fraction ($10^{-20}$) of the sun's mass, 
however all concentrated in a thin filament with a thickness of less than $0.004\,\mathrm{fm}$.

The results presented here are qualitatively similar to those from previous investigations 
that instead of featuring a twisted Higgs field allowed for a non-trivial gauge-field 
admixture in the cosmic string~\cite{Graham:2011fw} as variants of the Nielson-Olesen 
configuration~\cite{Nielsen:1973cs}; the gauge field component of the optimal
configuration turned out to be marginal. In fact, the presently obtained fermion mass 
and charges necessary to stabilize a string are only about $10\%$ larger than those in 
the previous study. This indicates that the dominant mechanism in the binding of the 
cosmic string, {\it i.e.} the attraction of fermions, is mainly due to the small 
Higgs \emph{vev} seen by fermions that are strongly bound in the vicinity of the string. 
Complicated gauge field additions or topological windings play, apparently, a minor role.

The results presented in this work are interesting in their own right, as they 
show that a potential fourth generation of heavy quarks (with masses 
$m > 320\,\mathrm{GeV}$) that couple to the Higgs condensate in the standard way
can exist neither today nor in the early universe (in sufficient numbers) without 
causing the generation of stable cosmic strings that eventually form networks. 
Such networks would be detectable {\it e.g.}~by their gravitational lensing 
or their distortion of the cosmic microwave background, and can therefore be ruled out by 
experiment. Although our reasoning was made in a simplified version of the standard model, 
we believe that the qualitative effect carries over to the full electroweak theory
since enlarging the variational space can only lower the energy.

The simplified configuration of a bound string achieved in the present work allows to 
study extended networks of realistic cosmic strings in a more accessible framework in 
which fermions couple to a prescribed Higgs background without dynamical gauge fields.

Nevertheless, the hedgehog string configuration for the Higgs field can be augmented 
by a gauge field component. Adopting Weyl gauge the decomposition of a possible hedgehog
gauge field must have the same structure as $\Phi^\dagger\Vek{\nabla}\Phi$ from 
Eq.~(\ref{config}),
\begin{equation}
\Vek{W}(\Vek{r})=\hat{\Vek{r}}\begin{pmatrix}
A(r) & i{\rm e}^{i\varphi}\, B(r) \cr
i{\rm e}^{-i\varphi}\, B(r) & A(r)\end{pmatrix}
+\frac{i}{r}\hat{\Vek{\varphi}}\begin{pmatrix}
a(r) & -i{\rm e}^{i\varphi}\, b(r) \cr
i{\rm e}^{i\varphi}\, b(r) & -a(r)\end{pmatrix}\,,
\label{gaugehedge}
\end{equation}
which introduces up to four additional radial functions in the plane perpendicular to 
the string; all of which vanish asymptotically. Of course, this expands the variational 
computation significantly. As a first simplification, the Higgs configuration would be 
fixed at the optimal configuration established in the current study.

\acknowledgments
N.~G.\ is supported in part by the NSF through grant PHY-1520293.
H.~W.\ is supported in part by the NRF (South Africa) by grant~109497.
\appendix
\section{Scattering off a hedgehog type of string}
\label{app:scat}
We solve the multi-channel scattering problem of a Dirac fermion in $D=2+1$ dimensions 
subject to the single particle Hamiltonian, Eq.~(\ref{Ham}). We employ planar polar 
coordinates $(r,\varphi)$ and perform a partial wave decomposition. Since neither the 
$z$-component of the nor the total angular momentum
$J_z = L_z + S_z$ nor isospin $I_z$ are separately conserved, and we label the solutions 
of the free Dirac equation by the eigenvalue $G \in \mathbb{Z}$ of the \emph{grand spin} 
operator $G_z = J_z + I_z$. The quantum number $\ell\in\mathbb{Z}$ of $L_z$ is determined 
by the angular dependency $e^{i \ell \varphi}$. For each value of $\ell$ there are four 
solutions of the free Dirac equation with given energy $\epsilon$ (and 4 solutions with 
energy $-\epsilon$ related by charge conjugation). These degenerate solutions do not all 
have the same angular dependence, since the free Hamiltonian contains $\varphi$-dependent 
terms and $\ell$ is not a good quantum number. However, we can still use it as an angular 
momentum \emph{channel index} in the partial wave decomposition. The actual angular 
dependence of the four degenerate solutions to the free Dirac equation is
\begin{align}
\langle \varphi \vert (\ell+1) + + \rangle &= e^{i(\ell+1) \varphi}\,
\begin{pmatrix} 1 \\ 0 \end{pmatrix}_S \otimes \begin{pmatrix} 1 \\ 0 \end{pmatrix}_I
\nonumber \\[2mm]
\langle \varphi \vert \ell + - \rangle &= (-i) e^{i\ell \varphi}\,
\begin{pmatrix} 1 \\ 0 \end{pmatrix}_S \otimes \begin{pmatrix} 0 \\ 1 \end{pmatrix}_I
\nonumber \\[2mm]
\langle \varphi \vert (\ell+2) - + \rangle &= i\,e^{i(\ell+2) \varphi}\,
\begin{pmatrix} 0 \\ 1 \end{pmatrix}_S \otimes \begin{pmatrix} 1 \\ 0 \end{pmatrix}_I
\nonumber \\[2mm]
\langle \varphi \vert (\ell+1) - - \rangle &= e^{i(\ell+1) \varphi}\,
\begin{pmatrix} 0 \\ 1 \end{pmatrix}_S \otimes \begin{pmatrix} 0 \\ 1 \end{pmatrix}_I \,.
\label{angspin}
\end{align}
The subscripts $S$ and $I$ indicate that the corresponding two-component spinors dwell in 
spin and isospin spaces, respectively. Each of these solutions is then considered as a 
four-component angular spinor. These states have grand spin $G=\ell$ or $G=\ell+2$, 
respectively, and this quantum number is conserved by the free Hamiltonian. 
The channel index $\ell \in \mathbb{Z}$ is signed, but channels $\ell$ and $-(\ell+2)$ are 
related by symmetry, so that we can restrict $\ell = -1, 0, 1, 2,\ldots$ with 
degeneracy $D_\ell = 2 - \delta_{\ell,-1}$. 

From the set of spinors in Eq.~(\ref{angspin}) we always combine those
with equal grand spin and dress them by radial functions to establish
the basis of the partial wave decomposition, 

\begin{align}
\psi_1(r,\varphi) &= \begin{pmatrix} 
                    f_1(r) \langle \varphi \vert (\ell+1) + + \rangle \\
                    g_1(r) \langle \varphi \vert (\ell+2) - + \rangle 
                    \end{pmatrix}
\qquad \qquad \quad G = \ell + 2
\nonumber \\[2mm]
\psi_2(r,\varphi) &= \begin{pmatrix} 
                    f_2(r) \langle \varphi \vert (\ell+0) + - \rangle \\
                    g_2(r) \langle \varphi \vert (\ell+1) - - \rangle 
                    \end{pmatrix}
\qquad \qquad \quad G = \ell
\nonumber \\[2mm]
\psi_3(r,\varphi) &= \begin{pmatrix} 
                    f_3(r) \langle \varphi \vert (\ell+2) - + \rangle \\
                    g_3(r) \langle \varphi \vert (\ell+1) + + \rangle 
                    \end{pmatrix}
\qquad \qquad \quad G = \ell + 2
\nonumber \\[2mm]
\psi_4(r,\varphi) &= \begin{pmatrix} 
                    f_4(r) \langle \varphi \vert (\ell+1) - - \rangle \\
                    g_4(r) \langle \varphi \vert (\ell+0) + - \rangle 
                    \end{pmatrix}
\qquad \qquad \quad G = \ell\,.
\label{app:basis}
\end{align}
Each of these eight-component spinors is a regular solution to the free Dirac 
equation when
\begin{align}
f^{(0)}_i(r) = J_\alpha(k r)\qquad {\rm and}
\qquad g^{(0)}_i(r) = \frac{\epsilon-m}{k}\,J_\beta(k r) \,,
\label{freereg}
\end{align}
where $|\epsilon| \ge m$ with $k = \sqrt{\epsilon^2 - m^2} > 0$.
The order of the Bessel function is determined by the angular momentum
associated with radial function, {\it i.e.} for $i=3$ we have 
$\alpha=\ell+2$ and $\beta=\ell+1$.

When the interaction $H_{\rm int}$ in Eq.~(\ref{hint}) is switched
on, the radial functions differ from the free case \eq{freereg} and mix 
among each other. To compactly formulate the resulting scattering problem 
we define two-component objects
\begin{equation}
\vec{u}(r) = \begin{pmatrix}f_1(r) \\ f_4(r) \end{pmatrix}\,,
\quad \vec{v}(r) = \begin{pmatrix}g_1(r) \\ g_4(r) \end{pmatrix}\,,
\quad \vec{w}(r) = \begin{pmatrix}f_2(r) \\ f_3(r) \end{pmatrix}
\quad {\rm and} \quad 
\vec{h}(r) = \begin{pmatrix}g_2(r) \\ g_3(r) \end{pmatrix}\,.
\label{110}
\end{equation}
The Dirac equation reduces to two sets of ordinary differential equations (ODE)
\begin{align}
 (\epsilon-m)\,\vec{u} &=  \mathbf{D}\,\vec{v} - \mathbf{X}\cdot \vec{u} 
 + \mathbf{Y}\cdot \vec{v} 
 \nonumber\\[2mm]
 (\epsilon+m)\,\vec{v} &= \overline{\mathbf{D}}\,\vec{u}  + \mathbf{X}\cdot \vec{v} 
 -\mathbf{Y}\cdot \vec{u}
 \label{ode1}
\end{align}
for $\vec{u}$ and $\vec{v}$ and
\begin{align}
 (\epsilon-m)\,\vec{w} &= \widehat{\mathbf{D}}\,\vec{h} - \mathbf{X}\cdot \vec{w} 
 - \mathbf{Y}\cdot \vec{h}
 \nonumber\\[2mm]
 (\epsilon+m)\,\vec{h} &=  \widehat{\overline{\mathbf{D}}}\,\vec{w} 
 + \mathbf{X}\cdot \vec{h} + \mathbf{Y}\cdot \vec{w}
 \label{ode2}
\end{align}
for $\vec{w}$ and $\vec{h}$. The separation into two decoupled sets is a feature of the 
hedgehog configuration, Eq.~(\ref{config}) and does not occur when gauge fields are 
included \cite{Graham:2011fw}. The boldface objects are $2\times2$ matrix operators. The 
radial derivatives and the centrifugal barriers are combined in the diagonal matrices
\begin{alignat}{3}
\mathbf{D} &= \text{diag}\left(
\frac{\ell+2}{r} + \partial_r \,,\, \frac{\ell}{r} - \partial_r \right)
\qquad&\qquad
\overline{\mathbf{D}} &=\text{diag}\left(
\frac{\ell+1}{r} - \partial_r \,,\, \frac{\ell+1}{r} + \partial_r \right)
\\[3mm]
\widehat{\mathbf{D}} &=\text{diag}\left(
\frac{\ell+1}{r} + \partial_r \,,\, \frac{\ell+1}{r} - \partial_r \right)
&
\widehat{\overline{\mathbf{D}}} &=\text{diag}\left(
\frac{\ell}{r} - \partial_r \,,\,\frac{\ell+2}{r} + \partial_r \right)\,.
\end{alignat}
The interaction matrices are expressed in terms of the profile
functions in Eq.~(\ref{profiles}),
\begin{align}
\mathbf{X} =m
\begin{pmatrix}
1- s(r) & 0 \\ 0 &  1 - s(r)  
\end{pmatrix} 
\qquad\qquad
\mathbf{Y} =m
\begin{pmatrix}
0 & p(r) \\ - p(r) & 0          
\end{pmatrix}\,.
\end{align}
For given energy $|\epsilon| >m$ and angular momentum $\ell$ we identify outgoing free polar 
waves, which are parameterized by Hankel functions of the first kind $H_\nu^{(1)}(kr)$.
We concentrate on the system eq.~(\ref{ode1}); the second system eq.~(\ref{ode2}) can be 
treated analogously. In the free case, the two linear independent complex polar wave solutions for 
$\vec{u\,}^{(0)}$ and $\vec{v\,}^{(0)}$ can be conveniently placed into the columns of 
two $2\times2$ matrices,
\begin{equation}
\mathcal{H}_u = \mathrm{diag}\Big(\,H^{(1)}_{\ell+1}(kr),\,H^{(1)}_{\ell+1}(kr)\,\Big)
\qquad {\rm and}\qquad
\mathcal{H}_v= \kappa\cdot \mathrm{diag}\Big(\,H^{(1)}_{\ell+2}(kr),\,
H^{(1)}_{\ell}(kr)\,\Big)\,,
\end{equation}
where
\begin{align}
\kappa = \frac{k}{\epsilon+m} = \frac{\epsilon-m}{k} \,.
\label{appkappa}
\end{align}
It is important to parameterize $\kappa$ as an odd function of $k$
because although $\kappa =\sqrt{\frac{\epsilon-m}{\epsilon+m}}$ is correct 
for $k\ge0$, it is deceptive for analytic continuation. Similarly we put the two linearly 
independent solutions of the full ODE system (\ref{ode1}) for $\vec{u}=(f_1,f_4)$ and 
$\vec{v}=(g_1,g_4)$ into the columns of $2\times2$ matrices
\begin{align}
\mathbf{U} = \Big( \vec{u}^{(1)}(r), \,\vec{u}^{(2)}(r)\Big)
\qquad{\rm and}\qquad 
\mathbf{V} = \Big( \vec{v}^{(1)}(r),\, \vec{v}^{(2)}(r)\Big)\,, 
\end{align}
respectively. It is convenient to factor out the free part and define
\begin{align}
\mathbf{U} = \mathcal{F}\cdot \mathcal{H}_u 
\qquad\qquad
\mathbf{V} = \mathcal{G}\cdot \mathcal{H}_v\,,
\end{align}
where the new Jost matrices obey the boundary conditions
\begin{align}
\lim_{r \to \infty} \mathcal{F}(r) = \lim_{r \to \infty} \mathcal{G}(r) = \ID\,. 
\end{align}
Inserting these ans{\"a}tze Eq.~(\ref{ode1})
yields the following equations for the $2\times2$ Jost matrices
\begin{align}
\partial_r \mathcal{F} &= \Big[\mathbf{\Lambda}_F - \mathbf{C}\mathbf{Y}\Big]\,\mathcal{F}
+ \mathcal{F}\,\Big[ + k \,\mathbf{C}\,\mathbf{Z}_F - \mathbf{\Lambda}_F\Big] + 
\Big[ - k \,\mathbf{C} + \kappa \,\mathbf{C}\mathbf{X}\Big]\,\mathcal{G}\,\mathbf{Z}_F
\nonumber \\[2mm]
\partial_r \mathcal{G} &= \Big[\mathbf{\Lambda}_G - \mathbf{C}\mathbf{Y}\Big]\,\mathcal{G}
+ \mathcal{G}\,\Big[ - k \,\mathbf{C}\,\mathbf{Z}_G - \mathbf{\Lambda}_G\Big] + 
\Big[ + k \,\mathbf{C} + \frac{1}{\kappa} \,\mathbf{C}\mathbf{X}\Big]\,\mathcal{F}\,\mathbf{Z}_G\,.
\label{802}
\end{align}
Here, the $2\times2$ matrix $\mathbf{C}={\rm diag}(1 ,-1)$ inverts the sign of the lower component.
The Hankel functions and centrifugal terms, which are of kinematic origin, enter through the matrices
\begin{alignat}{3}
\mathbf{Z}_F &= \text{diag}\,\left( 
\frac{H_{\ell+2}^{(1)}(k r)}{H_{\ell+1}^{(1)}(k r)}\,,\,
\frac{H_{\ell}^{(1)}(k r)}{H_{\ell+1}^{(1)}(k r)}\right)\,,
\qquad&\qquad
\mathbf{Z}_G &= 
\text{diag}\,\left( \frac{H_{\ell+1}^{(1)}(k r)}{H_{\ell+2}^{(1)}(k r)}\,,\,
\frac{H_{\ell+1}^{(1)}(k r)}{H_{\ell}^{(1)}(k r)} \right)\,,
\nonumber \\[3mm]
\mathbf{\Lambda}_F &=  \frac{1}{r}\,\text{diag}\,\left( 
\ell+1\,,\,-(\ell+1)\right)\,,
&
\mathbf{\Lambda}_G &=  \frac{1}{r}\,\text{diag}\,\left( 
-(\ell+2)\,,\,\ell\right)\,.
\label{x28}
\end{alignat}
We observe that asymptotically, {\it i.e.} $r\to\infty$, the first columns of $\mathbf{U}$ and 
$\mathbf{V}$ correspond to an outgoing wave only in the channel $\psi_1$ while the second 
columns have an outgoing wave only in the channel $\psi_4$. Finally noting that the complex 
conjugate of the Jost solution also solves the (real) radial ODE system the scattering wave 
function is the linear combination
\begin{align}
\Psi_u = \mathcal{F}^\ast\cdot \mathcal{H}_u^\ast + \mathcal{F}\cdot \mathcal{H}_u 
\cdot \mathcal{S}\,.
\end{align}
The $S$-matrix is determined by the requirement that $\Psi_u$ is 
regular at the origin $r\to0$, with the result 
\begin{align}
\mathcal{S} = - \lim_{r \to 0}\mathcal{H}_u^{-1}
\cdot \mathcal{F}^{-1}\cdot \mathcal{F}^\ast \cdot \mathcal{H}_u^\ast 
= - \lim_{r \to 0}\mathcal{H}_v^{-1}
\cdot \mathcal{G}^{-1}\cdot \mathcal{G}^\ast \cdot \mathcal{H}_v^\ast\,.
\label{824}
\end{align}

As mentioned in the main text, it is advantageous to find the Jost matrix for momenta 
analytically continued to the imaginary axis, $k\to it$ with $t>0$, since the 
resulting spectral integral, Eq.~(\ref{evac0}) fully accounts for the bound state 
contribution to $\mathcal{E}_q$. The continuation must, in principle, be carried 
out separately for both signs of the energy $\epsilon = \pm \sqrt{m^2 + k^2}$. 
In the present case, the theory is charge-conjugation invariant for real momenta
and we can select one sign of the energy (say, $\epsilon > 0$). The second Riemann 
sheet then contributes an overall factor of two to the vacuum energy per unit length, 
{\it cf}.~Eq.(\ref{evac2}).  For simplicity, we only present the derivation for Eq.~(\ref{ode1}); 
the corresponding results for Eq.~(\ref{ode2}) can be obtained by some simple sign changes 
and angular momentum relabelings.

If we assume that the Jost matrices $\mathcal{F}$ and $\mathcal{G}$, Eq.~(\ref{802})
are analytic functions of the momentum, the naive continuation $k \to it$ yields
\begin{align}
\partial_r \mathcal{F} &= \Big[\mathbf{\Lambda}_F - \mathbf{C}\mathbf{Y}\Big]\,\mathcal{F}
+ \mathcal{F}\,\Big[ t \,\mathbf{C}\,\mathscr{Z}_F - \mathbf{\Lambda}_F\Big] + 
\Big[ - t \,\mathbf{C} + z_k^\ast \,\mathbf{C}\mathbf{X}\Big]\,\mathcal{G}\,\mathscr{Z}_F
\nonumber \\[2mm]
\partial_r \mathcal{G} &= \Big[\mathbf{\Lambda}_G - \mathbf{C}\mathbf{Y}\Big]\,\mathcal{G}
+ \mathcal{G}\,\Big[ - t \,\mathbf{C}\,\mathscr{Z}_G - \mathbf{\Lambda}_G\Big] + 
\Big[ t \,\mathbf{C} - z_k \,\mathbf{C}\mathbf{X}\Big]\,\mathcal{F}\,\mathscr{Z}_G\,.
\label{1002} 
\end{align}
Here, $\mathcal{F}=\mathcal{F}(it,r)$ and $\mathcal{G}=\mathcal{G}(it,r)$ are again complex $2\times2$ 
matrices. The kinematical factor $\kappa$ from Eq.~(\ref{appkappa}) has turned into a pure phase
\begin{align}
\kappa \stackrel{k \to it}{\longrightarrow} i\,z_k^\ast\,,\qquad\qquad
z_k = \frac{m + i \sqrt{t^2 - m^2}}{t} = \frac{1}{z_k^\ast}
\end{align}
and the Hankel functions are replaced by modified Bessel functions contained in
\begin{align}
\mathscr{Z}_F &\equiv i \mathbf{Z}_F(it r) = \mathrm{diag}
\,\left( \frac{K_{\ell+2}(t r)}{K_{\ell+1}(t r )}\,,\,- \frac{K_{\ell}(t r)}{K_{\ell+1}(t r )}
\right)
\nonumber \\[2mm]
\mathscr{Z}_G &\equiv i \mathbf{Z}_G(it r) = \mathrm{diag}
\,\left( - \frac{K_{\ell+1}(t r)}{K_{\ell+2}(t r )}\,,\,\frac{K_{\ell+1}(t r)}{K_{\ell}(t r )}
\right)\,.
\label{x42}
\end{align}
The Born series is obtained by expanding these differential equations in powers of the interaction. 
The leading term is always the $2\times2$ unit matrix, so that 
$\mathcal{F} = \ID + \mathcal{F}_1 + \mathcal{F}_2+ \ldots$ and
$\mathcal{G} = \ID + \mathcal{G}_1 + \mathcal{G}_2+ \ldots$. 
This expansion leads to 
\begin{align}
\partial_r \mathcal{F}_1 &=  \big[\mathbf{\Lambda}_F \,,\,\mathcal{F}_1\big]
+ t \,\big(\mathcal{F}_1\,\mathbf{C} - \mathbf{C}\,\mathcal{G}_1\big)\,\mathscr{Z}_F
+ z_k^\ast \,\mathbf{C}\mathbf{X}\,\mathscr{Z}_F - \mathbf{C}\mathbf{Y}
\nonumber \\[2mm]
\partial_r \mathcal{G}_1 &=  \big[\mathbf{\Lambda}_G \,,\,\mathcal{G}_1\big]
+ t \,\big(\mathbf{C}\,\mathcal{F}_1 - \mathcal{G}_1\,\mathbf{C}\big)\,\mathscr{Z}_G
- z_k \,\mathbf{C}\mathbf{X}\,\mathscr{Z}_G - \mathbf{C}\mathbf{Y}
\nonumber \\[2mm]
\partial_r \mathcal{F}_2 &=  \big[\mathbf{\Lambda}_F \,,\,\mathcal{F}_2\big]
+ t \,\big(\mathcal{F}_2\,\mathbf{C} - \mathbf{C}\,\mathcal{G}_2\big)\,\mathscr{Z}_F
+ z_k^\ast \,\mathbf{C}\mathbf{X}\,\mathcal{G}_1\,\mathscr{Z}_F 
- \mathbf{C}\mathbf{Y}\,\mathcal{F}_1
\nonumber \\[2mm]
\partial_r \mathcal{G}_2 &=  \big[\mathbf{\Lambda}_G \,,\,\mathcal{G}_2\big]
+ t \,\big(\mathbf{C}\,\mathcal{F}_2 - \mathcal{G}_2\,\mathbf{C}\big)\,\mathscr{Z}_G
- z_k \,\mathbf{C}\mathbf{X}\,\mathcal{F}_1\,\mathscr{Z}_G 
- \mathbf{C}\mathbf{Y}\,\mathcal{G}_1\,.
\label{1002-born} 
\end{align}

For the quantum energy we require the logarithmic Jost functions $\widetilde{\nu}(t)$ 
defined by
\begin{equation}
\exp\left[ \widetilde{\nu}_F(t)\right] = \lim_{r \to 0}\,\mathrm{det}\,\mathcal{F}(it,r)  
\qquad \mbox{and}\qquad
\exp\left[ \widetilde{\nu}_G(t)\right] = \lim_{r \to 0}\,\mathrm{det}\,\mathcal{G}(it,r)\,.
\label{logjost}
\end{equation}
These quantities have the Born expansion
\begin{align}
\widetilde{\nu}_F(t) &= \mathrm{tr}\,\mathcal{F}_1  + \mathrm{tr}\,\Big( \mathcal{F}_2 - 
\frac{1}{2}\,\mathcal{F}_1\cdot \mathcal{F}_1\Big) + \cdots
\equiv \widetilde{\nu}_F^{(1)}(t) + \widetilde{\nu}_F^{(2)}(t) + \cdots
\nonumber \\[2mm]
\widetilde{\nu}_G(t) &= \mathrm{tr}\,\mathcal{G}_1  + \mathrm{tr}\,\Big( \mathcal{G}_2 - 
\frac{1}{2}\,\mathcal{G}_1\cdot \mathcal{G}_1\Big) + \cdots
\equiv \widetilde{\nu}_G^{(1)}(t) + \widetilde{\nu}_G^{(2)}(t) + \cdots\,.
\label{fuzz}
\end{align}
To find the relationship between $\widetilde{\nu}_F(t)$ and $\widetilde{\nu}_G(t)$ and, 
most importantly $\nu(t)=\nu_\ell(t)$ that enters Eq.~(\ref{defu}), we recall
that the Jost function is defined by the Wronskian between the Jost solution and the 
\emph{regular} solution. The latter is defined by a momentum-independent boundary condition 
at the origin $r\to0$. As $r\to0$ the Higgs field does not assume its \emph{vev}, {\it i.e.} 
$s(0)\ne 1$, {\it cf.} Eq.~(\ref{profiles}). This changes the kinematical quantities in 
Eq.~(\ref{freereg}) of the regular solution to
\begin{align}
f^{(0)}_i(r) = J_\alpha(q r)\qquad {\rm and}
\qquad g^{(0)}_i(r) = \zeta\,J_\beta(q r) \,,
\label{freereg1}
\end{align}
where
\begin{align}
q^2 = \epsilon^2 - m^2s^2(0)\qquad{\rm and}\qquad 
\zeta^2 = \frac{\epsilon-ms(0)}{\epsilon + ms(0)} \,.
\end{align}
Working out the Wronskian yields the following 
correction for the logarithmic Jost function and its Born series \cite{Graham:2011fw},
\begin{align}
\nu_F(t) &\equiv \widetilde{\nu}_F(t) + 2\, \ln\left(\frac{\tau - im}{\tau - ims(0)}\right)
& 
\nu_G(t) &\equiv \widetilde{\nu}_G(t) + 2\, \ln\left(\frac{\tau + im}{\tau + ims(0)}\right)
\nonumber \\[2mm]
\nu_F^{(1)}(t) &\equiv \widetilde{\nu}^{(1)}_F(t) + 2\, \frac{1-s(0)}{1 + i\,\tau/m}
& 
\nu_G^{(1)}(t) &\equiv \widetilde{\nu}^{(1)}_G(t) +2 \, \frac{1-s(0)}{1 - i\,\tau/m}
\nonumber \\[2mm]
\nu_F^{(2)}(t) &\equiv\widetilde{\nu}^{(2)}_F(t) + \left(
\frac{1 - s(0)}{1 + i \,\tau/m}\right)^2
& 
\nu_G^{(2)}(t) &\equiv\widetilde{\nu}^{(2)}_G(t) + \left(
\frac{1 - s(0)}{1 - i \,\tau/m}\right)^2\,,
\label{zruf}
\end{align}
where $\tau = \sqrt{t^2-m^2}$ and the factor two arises because there are
four channels: $\ln \zeta^4=2\ln\zeta^2$. 

With these modifications we find that $\nu_F(t)$ and $\nu_G(t)$ are indeed real and identical.
This is also true at any order in the Born series. The pseudo-scalar profile component does 
not contribute to the correction because $p(0)=0$. 

\section{Feynman diagrams}
\label{app:feyn}
The Feynman diagrams are generated by the expansion of the fermion determinant
\begin{align}
\mathcal{A} \equiv - T L_z \,\mathcal{E}_F 
& = (-i)\,N_c\,\ln \mathrm{det}\,\left( - \partial\!\!\!/ - m - V\right) 
\nonumber \\[2mm]
& = (-i)\,N_c\,\ln \mathrm{det}\,(\partial\!\!\!/ -m) + 
iN_c \sum_{n=1}^\infty \frac{1}{n} 
\,\mathrm{Tr}\,\Big[ \big(i \partial\!\!\!/ - m\big)^{-1}\,V\Big]^n 
\equiv \sum_{n=0}^\infty \mathcal{A}_n\,,
\end{align}
where $V=\beta H_{\rm int}$ is the interaction potential from Eq.~(\ref{hint}).
The first-order ($n=1$) diagram is \emph{local} and can be eliminated completely 
by a counterterm of the form
\[
\mathcal{L}_{CT} = c_3\,\Big[\mathrm{tr}(\Phi^\dagger\Phi) - 2 v^2\Big]
\]
which contains $s(r)-1$, $\left[s(r)-1\right]^2$ and $p^2(r)$ terms. The linear term 
eliminates the tadpole and keeps the Higgs \emph{vev} at its classical value. The 
quadratic terms serve to renormalize $\mathcal{A}_2$, together with the quadratic part 
of the second counterterm
\begin{align}
\mathcal{L}_{CT} = c_4\,\Big[\mathrm{tr}(\Phi^\dagger\Phi) - 2 v^2\Big]^2\,.
\label{c4ct}
\end{align}
This counterterm also contains pieces cubic and quartic in the 
profiles. They renormalize the third- and fourth-order diagram below. Choosing the 
no-tadpole scheme for $\mathcal{A}_1$ and $\overline{MS}$ for $\mathcal{A}_2$ yields
\begin{align} \mathcal{E}_{\rm FD}^{(2)}\Big|_{\overline{MS}} 
\equiv \frac{-1}{TL_z}\,\big[\mathcal{A}_1 
+ \mathcal{A}_2\big] = -N_c\, \int_0^\infty \frac{dk\,k}{4 \pi}\,I_1(k/m)\,
\Big( 4 m^2 \, \widetilde{\alpha}_H(k)^2 + k^2\,\big[\widetilde{\alpha}_H(k)^2 
+ \widetilde{\alpha}_P(k)^2 \big] \Big) 
\label{app_Efermi}
\end{align}
with the explicit parameter integral 
\begin{align} I_1(t) &\equiv \int_0^1 dx\,\ln\big[1 + x (1-x)\,t^2\big] 
= \frac{2}{t}\,\sqrt{4 + t^2}\, \mathrm{arcsinh}(t/2) - 2
\label{eqn:I1}
\end{align}
and the Fourier-Bessel transform of the background potential
\begin{align} 
\widetilde{\alpha}_H(k) &= m \int_0^\infty dr\,r\,J_0(k r )\,
\big[ s(r) - 1 \big] 
\\[2mm] 
\widetilde{\alpha}_P(k) &= m \int_0^\infty dr\,r\,J_1(kr)\,p(r)\,. 
\label{alphamom}
\end{align}
The contribution quadratic in $\widetilde{\alpha}_P(k)$ starts with a prefactor 
$k^2$, {\it i.e.}~the pseudo-scalar excitations remain massless.

\medskip\noindent
The third- and fourth-order diagrams are more complicated. Fortunately, within the fake boson 
method, {\it cf.} the following appendix, we only need to identify their (logarithmic) divergences
\begin{align}
\big[\mathcal{A}_3 + \mathcal{A}_4\big] 
= i\,\pi\,c_F\, T L_z \,\mu^{4-D} \int \frac{d^D k}{(2\pi)^D}\,
(k^2 - m^2 + i 0)^{-2} + \ldots\,,
\label{appdiv}
\end{align}
where D is the number of spacetime dimensions in dimensional regularization and the ellipsis
indicate finite pieces. Since the only counterterm for these diagrams is Eq.~(\ref{c4ct})
and the coefficient $c_4$ has already been determined by the second-order diagram
above, we can predict $c_F$ directly if we assume that the theory is renormalizable.
Alternatively, we can compute $c_F$ from the divergence of the third- and fourth-order 
diagram, which yields the same expression 
\begin{align} 
c_F = 4m^4 N_c\int_0^\infty dr\,r\,\Big[\left(s(r)-1\right)^2+ p^2(r)\Big]
\left[\left(s(r)-1\right)^2 + p^2(r) + 4 s(r)-4\right]\, 
\label{appc3}
\end{align}
where the prefactor four results from the Dirac trace.

\section{Fake boson subtraction}
\label{app:fake}
The second-order Feynman diagram of a scalar boson scattering off a radially symmetric 
background potential $V_B(r)$ is logarithmically divergent. By proper rescaling it replaces 
the third- and fourth-order fermion diagrams and Born subtractions. To be specific, we 
choose a one-parameter profile 
\begin{align} 
V_B(r) \equiv m^2\frac{r}{w_B}\exp\left(-2 \frac{r}{w_B}\right)\,, 
\end{align}
where $w_B$ is an arbitrary width which should not play a role in the final result. 
The logarithmic divergence of the second-order contribution to the effective action 
\begin{align} 
\mathcal{A}_2^{(\infty)} &= i\,\pi \, c_B\, T L \mu^{4-D}\, 
\int \frac{d^D q}{(2\pi)^D}\,(q^2 - m^2 + i 0)^{-2} 
\nonumber \\[2mm] 
c_B &\equiv - \frac{1}{2}\int_0^\infty dr\,r \,V_B(r)^2 
= - \frac{3m^4\, w_B^2}{256}\,, 
\end{align}
where $\mu$ is an arbitrary renormalization scale introduced by dimensional regularization 
to $D$ spacetime dimensions. This should be compared to the corresponding expression 
Eq.~(\ref{appdiv}) from the third- and fourth-order
fermion diagrams. Employing the
$\overline{MS}$ scheme, the renormalized energy per unit length is
\begin{align} 
\mathcal{E}_{B}^{(2)}\Big|_{\overline{MS}} &= + \frac{1}{32 \pi} \int_0^\infty dq\,q 
I_1(q)\,\overline{V}_B(q)^2
\label{app_Efake}
\end{align}
where $I_1$ is given in Eq.~(\ref{eqn:I1}) and $q \equiv k/m$ is
dimensionless. The Fourier transform of the background is also dimensionless 
\begin{align} 
\overline{V}_B(q) \equiv \int_0^\infty dr\,r V_B(r)\,J_0(q m r) 
= (\widehat{w}_B)^2 \frac{8 - (\widehat{w}_B q)^2}
{\big[4 + (\widehat{w}_B q)^2 \big]^{\frac{5}{2}}}\,, 
\end{align}
where $\widehat{w}_B \equiv m w_B$ is the fake boson profile width measured 
in inverse units of the fermion mass $m$.

\medskip\noindent
The second-order Born approximation, $\overline{\nu}_\ell^{(2)}(k)$, to the logarithm of 
the Jost function for a scalar boson scattering off the background $V_B$ can be computed 
by standard techniques, {\it cf.} Ref.~\cite{Graham:2002xq}. After analytic continuation 
to the imaginary axis it gives  rise to the function 
\begin{align}
u_B^{(2)}(t) \equiv \sum_{\ell=0}^\infty \left[2 - \delta_{\ell 0}\right]\,
\overline{\nu}_\ell^{(2)}(it)\,, 
\label{defuB}
\end{align}
which enters Eq.~(\ref{bosinter}) and produces a finite spectral integral
in Eq.~(\ref{evac2}). Numerically we have verified invariance with respect to the 
artificial width parameter $w_B$.

\section{On-shell renormalization scheme}
\label{app:onshell}
All finite counterterm contributions contain pieces from the classical Lagrangian with finite 
coefficients,
\begin{align} 
\Delta \mathcal{E}_{\rm ren} &= N_c\,\int_0^\infty dr\,r\,\Bigg\{\overline{c}_2\,
\Big[s'(r)^2 + p'(r)^2 + \frac{p(r)^2}{r^2}\Big] + 
\overline{c}_4\,\Big[1 -s(r)^2 - p(r)^2\Big]^2\Bigg\}\,,
\label{app_Eren}
\end{align}
where the prime denotes the derivative with respect to the radial coordinate $r$.
When passing from the $\overline{MS}$ to the physical \emph{on-shell}
scheme, the finite coefficients $\overline{c}_2$ and $\overline{c}_4$ are determined such that the
renormalized Higgs propagator has a pole at $4\lambda v^2$ with unit residue.
The general expressions are readily taken from Ref. \cite{Graham:2011fw}. Fortunately
they simplify considerably for the case of the hedgehog string,
\begin{equation} 
\overline{c}_2 = \frac{1}{\pi}\,\Big[ \frac{1}{3}+3\,I_2(i\mu_H) \Big] 
\qquad {\rm and}\qquad
\overline{c}_4 = \frac{1}{4\pi}\,\Big[\mu_H^2 + 6 \, I_1(i \mu_H) \Big] 
\end{equation}
where $I_1(i \mu_H)$ is given in Eq.~(\ref{eqn:I1}),
$\mu_H = m_H / m = 2 \sqrt{\lambda} / f$, and 
\begin{align} 
I_2(i\mu) &= \int_0^1 dx\,x (1-x)\ln\big[1 - x(1-x)\,\mu^2\big] 
= - \frac{\mu (12 + 5 \mu^2) \,\sqrt{4 - \mu^2} + 6 (\mu^4 - 2 \mu^2 - 8) 
\,\arcsin(\mu/2)} {18 \mu^3\,\sqrt{4 - \mu^2}}\,.
\end{align}

\section{Bound states}
\label{app:charge}

In this appendix we describe the computation of the single particle bound state energies, 
$\epsilon_{i,\ell}$. We follow Ref.~\cite{Graham:2011fw} and diagonalize the interaction 
Hamiltonian, Eq.~(\ref{Ham}) in the free grand spin basis used in appendix \ref{app:scat},
{\it cf}. Eqs.~(\ref{app:basis}) and (\ref{freereg}).  The discretized momenta $k_n^{(\ell)}$ 
in the angular momentum channel $\ell$ are determined such that no flux emerges from the string 
core through a large circle of radius~$R$ around the core. The flux combines upper and lower 
components of the spinor in Eq.~(\ref{app:basis}) and vanishes when any of them is zero. From 
Eq.~(\ref{freereg}) it is obvious that the most compact condition is
\[
J_{\ell+1}(k_n^{(\ell)} R) = 0 \,,\qquad\quad n = 1,2,\ldots\,.
\]
Since for any given $\ell$ there is only one set of discretized momenta, we will omit that 
label for simplicity.

We impose a numerical cutoff $\Lambda$ such that only the $k_n<\Lambda$ are included in the basis. 
The total number $N$ of such momenta $k_n$ depends on both the angular momentum channel $\ell$ and 
the size of the radius~$R$.  For each momentum $k_n$ there are two, which we sort in ascending order: 
\begin{align}
 \epsilon^{(0)}_n = 
 \begin{cases}
 - \sqrt{k_{N+1-n}^2 - m^2} \quad &:\quad n = 1,\ldots,N 
 \\[2mm]
 + \sqrt{k_{n-N}^2- m^2} \quad &: \quad n = N+1,\ldots,2N\,.
 \end{cases}
 \label{eigenfree}
\end{align}
The free Hamiltonian, Eq.~(\ref{hfree}), exhibits a four-fold degeneracy 
from spin and isospin invariance, which we assemble into a single 
super-index that has two entries $\alpha=(n,i)$ with $i=1,2,3,4$, 
according to Eq.~(\ref{app:basis}).
The interaction matrix elements are worked out explicitly using the
super-indices $\alpha=(n,i)$ and $\beta=(m,j)$
\begin{align}
\hat{V}_{(n,i)(m,j)} = \langle n i \,\vert\, H_{\rm int}\,\vert\,m j \rangle & = 
\delta_{ij}\,\int_0^Rdr\,r\,\Big[f^{(0)}_i(k_n r)\, f^{(0)}_j(k_m r) 
- g^{(0)}_i(k_n r)\, g^{(0)}_j(k_m r)\Big]
\nonumber \\[2mm]
&\qquad + \sigma_{ij} \int_0^R dr\,r\,\Big[f^{(0)}_i(k_n r)\,g^{(0)}_j(k_m r) 
- g^{(0)}_i(k_n r)\, f^{(0)}_j(k_m r) \Big]\,,
\end{align}
where $f^{(0)}_i$ and $g^{(0)}_i$ are the radial functions from 
Eq.(\ref{freereg}) with momenta $k_n$ and
\renewcommand{\arraystretch}{0.8}
\[
\sigma_{ij} = 
\begin{pmatrix}
0  &  0 &  0 & +1 \\
0  &  0 & -1 & 0  \\
0  & +1 &  0 & 0  \\
-1 & 0  &  0 & 0
\end{pmatrix}\,.
\]
\renewcommand{\arraystretch}{1.0}

Numerical diagonalization of the symmetric $8N \times 8N$ matrix $\hat{H}=\hat{H}_0 + \hat{V}$
($\hat{H}_0$ is a diagonal matrix that contains four copies of $\epsilon_n^{(0)}$)
yields $8N$ eigenvalues $\epsilon_{(n,i)}$. Those with $|\epsilon_{(n,i)}|<m$ are stable against changes 
of sufficiently large $\Lambda$ or $R$ and are identified as the true bound state energies. The 
numerical tests in section \ref{sec:results} indicate that $\Lambda\approx10m$ and $R\approx80/m$, 
which corresponds to $N \approx 250$, can be considered sufficiently large for all contributing 
angular momentum channels $\ell$. In that case we have to diagonalize a $2000 \times 2000$ matrix 
in every angular momentum channel.

\end{document}